\documentclass[11pt, a4paper]{article}

\usepackage[pages=all, color=black, position={current page.south}, placement=bottom, scale=1, opacity=1, vshift=5mm]{background}
\SetBgContents{
} % copyright

\usepackage[margin=1in]{geometry} % full-width
\usepackage{graphicx,subcaption,lipsum}
\usepackage{amsmath}
\usepackage{amsthm}
\usepackage{amssymb}
\usepackage{bbm}
\usepackage[english]{babel}
\addto{\extrasenglish}{%
  
}
\usepackage{subcaption}
\usepackage{booktabs}
\usepackage{listings}
\usepackage{multirow}
\usepackage{url}

\usepackage{multirow}
\usepackage{array}

\usepackage{lipsum}
\usepackage{setspace}

\usepackage{booktabs}
\usepackage{adjustbox}

\usepackage[rightcaption]{sidecap}

\usepackage[utf8]{inputenc}
\usepackage{hyperref}
\hypersetup{
	unicode,
	colorlinks,
	breaklinks,
	urlcolor=blue, 
	linkcolor=blue, 
	pdfauthor={Author One, Author Two, Author Three},
	pdftitle={A simple article template},
	pdfsubject={A simple article template},
	pdfkeywords={article, template, simple},
	pdfproducer={LaTeX},
	pdfcreator={pdflatex}
}

\usepackage{hyperref}
\usepackage[numbers]{natbib}

\usepackage[nottoc]{tocbibind}

\usepackage{graphicx}
\theoremstyle{plain}

\theoremstyle{definition}

\usepackage{graphicx, color}
\graphicspath{{fig/}}
\usepackage{algorithm, algpseudocode} % use algorithm and algorithmicx for typesetting algo
\usepackage{mathrsfs} % for \mathscr command
\usepackage{authblk}

\title{Spatiotemporal Dynamics of Conflict Occurrence and Fatalities in Ethiopia: A Bayesian Model and Predictive Insights Using Event-level Data (1997 -- 2024)}

\author[1,4]{Yassin Tesfaw Abebe\thanks{ytesfaw@yahoo.com}}
\author[1]{Abdu Mohammed Seid \thanks{Corresponding author: \texttt{abdum442@yahoo.com}}}
\author[2]{Lassi Roininen\thanks{lassi.roininen@lut.fi}}
\author[3]{Mohammed Seid Ali\thanks{mohammedseid1997@gmail.com}}

\affil[1]{Bahir Dar University, Department of Mathematics, Bahir Dar, Ethiopia}
\affil[2]{LUT University, School of Engineering Sciences, Lappeenranta, Finland}
\affil[3]{Bahir Dar University, Department of Political Science \& International Studies, Bahir Dar, Ethiopia}
\affil[4]{Mekdela Amba University, Department of Mathematics, Tulu-Awulia, Ethiopia}

\begin{document}
%\onehalfspacing
%\pagecolor{yellow!15}
\date{}
\maketitle
	
\begin{abstract}
This study presents a spatiotemporal, dual Bayesian model designed to examine both the occurrence and the number of conflict fatalities using event-level data applied to over 28 years of conflict data from Ethiopia spanning 1997–2024, sourced from the Armed Conflict Location and Event Data (ACLED) project. Conflict-related fatalities are typically composed of two linked dimensions: the binary occurrence of fatalities and the count of fatalities when they occur. The model includes an additive fixed effect component to model the covariates, along with a random effect component to capture the spatiotemporal influences on both outcomes, while allowing for specific effects for each outcome. Covariates considered in the model include event types and season as categorical variables, proximity to cities and borders as nonlinear effects, and population as an offset term in the count model. A latent spatiotemporal process is used to account for common spatial and temporal influences on both outcomes. The spatial structure is modeled using a Mat\'ern field prior, and inference is carried out through Integrated Nested Laplace Approximation (INLA). The results demonstrate significant spatial clustering and temporal fluctuations in fatality risks, reinforcing the value of incorporating both dimensions for a more profound understanding and better prediction of the dynamics of conflict-related violence. For event types, findings showed that airstrikes, shelling, and attacks had the highest likelihood of fatality occurrence and had the highest impact on the likelihood of multiple fatality counts. Foreign, communal, and rebel actors derive the highest conflict fatalities, while protesters and rioters show markedly lower fatality risks, reflecting the distinct impact of actor types. The finding also revealed that there is a higher likelihood of multiple fatalities in the summer season. The study also conclude that proximity to international borders is a primary driver of high-intensity violence, while remoteness from urban centers significantly increases the probability of lower-intensity fatal events. These findings provide new insight into the dynamics of conflict violence and offer practical value for planning, policy, and resource allocation to better protect vulnerable communities.

\vspace{.25in}
\noindent\textbf{Keywords: Spatiotemporal Analysis, Conflict Fatalities, Bayesian Modeling, INLA, SPDE Approach, ACLED, Ethiopia} 
\end{abstract}

%\tableofcontents

\section{Introduction}\label{sec_intro}
Ethiopia, located at the strategic crossroads of the Horn of Africa, has faced recurrent waves of violent conflict rooted in a complex interplay of historical, political, and socioeconomic factors. The country’s modern political and geopolitical history has included transitions from imperial rule to military dictatorship and, more recently, to an ethnically based federal system. Each of these shifts has generated new forms of contention and realigned social and political fault lines \citep{geda2005conflict, etefa2019ethnicity}. Over time, struggles for control over land and resources, competition among political elites, and the politicization of ethnic identity, along with various forms of foreign interventions because of the country's geopolitical vulnerabilities following its strategic hydropolitical roles over the Nile and the Red Sea, have interacted to produce both localized and widespread outbreaks of violence. Persistent divisions of various sections of the society, poor democratic culture, and politicization of ethnicity have further amplified conflict, resulting in patterns that are highly dynamic \citep{siyum2021underlying, tepfenhart2013causes}.

The above socio-economic, political, and strategic geopolitical factors embedded in the country's political history, along with limitations of the current administrations of the Prosperity Party (PP), have made political stability nearly impossible in Ethiopia even after the post-2018 political reform but have instead been followed by increasing intensity and types of conflicts across different parts of the country. The Tigray war (November 2020 -- November 2022) emerged as one of the world’s deadliest contemporary civil wars; estimates indicate hundreds of thousands of fatalities, including extensive civilian casualties \cite{mcgowan2025conflict}. The humanitarian consequences were severe; more than 2.6 million people were internally displaced, and hundreds of thousands more fled across borders. Critical infrastructures, including hospitals, schools, and utilities, were destroyed \cite{gebrihet2025armed}. The violence subsequently spread to the Afar and Amhara region, resulting in millions more displaced, destruction of private properties and civilian infrastructures, serious violation of humanitarian law including repeated targeting of civilian populations, various forms of sexual violence, and mass killings of innocent civilians as an act of ethnic violence against the Amhara and Afar communities \citep{OHCHR2023}.

Almost immediately after the cessation of hostilities in Tigray following the Pretoria Accord in November 2022, a new conflict front erupted in the Amhara National Regional State, where the Fano militia, formerly allied with the federal government, launched a widespread insurgency in 2024. Heavy fighting with the Ethiopian National Defense Force has led to significant casualties on both sides, while civilians have suffered from extrajudicial killings and aerial drone strikes that have struck residential areas, schools, and health centers. This renewed violence has displaced more than two million people in the Amhara region alone, overwhelming food systems, collapsing healthcare, and disrupting education for millions of children \citep{OHCHR2023, HRW2025}. Simultaneously, persistent communal violence in the Oromia, Somali, Afar, and Benishangul-Gumuz regions has displaced millions more, contributing to an ongoing, nationwide humanitarian emergency \cite{ACLED2024EPO}. These widespread crises are compounded by the weakness of state institutions and state fragility, corruption, ethnic politicization, and interference by neighboring states and non-state actors, all of which create opportunities for armed groups to aggravate the fragility of the Ethiopian state \cite{muhyie2025synthesizing}.

The cumulative impact of these concurrent conflicts has resulted in a humanitarian crisis of a wider scale. By 2024, over 4 million Ethiopians required urgent humanitarian assistance as attacks on health and educational facilities became commonplace, crippling access to basic government services. Agricultural production, the mainstay of most livelihoods, has been severely disrupted, raising the specter of famine. The educational crisis is acute: 4.4 million children are out of school in Amhara and Oromia alone \citep{UNICEF2025}. Moreover, violent conflict has a profound impact on food security, health, and economic stability. A study \citet{abay2023near} found that seven months into the conflict, the probability of moderate to severe food insecurity increased by 37 percent, with each additional battle exposure leading to a 1\%  rise in this probability. Another study in conflict and health reported that 57\% of households in war-torn Tigray experienced food insecurity, with 30.47\% mildly, 27.21\% moderately, and 21.86\% severely food-insecure \cite{gebrihet2025armed}. The study \citet{arage2023exploring} reveals that the 2022 conflict in Northeast Ethiopia inflicted both direct and indirect health harms ranging from casualties, displacement, and violence to systemic collapse of healthcare services and long-term issues like PTSD and disability. Economically, the conflict has led to a significant reduction in tax revenues and an increase in military expenditures, straining the nation's fiscal resources and hindering development efforts \citep{muriuki2023impact}. 
Recent studies by \citet{muhyie2025synthesizing} shows that armed conflict in the Amhara region devastated food systems, livelihoods, and social cohesion, displacing over 5.5 million people, destroying agricultural productivity, and causing infrastructure losses estimated at US \$500 million. Additionally, \citet{biset2023effect} finds that conflict in the region caused severe displacement (35\%), high disease burden (41\%), widespread violence (70\%), and acute malnutrition (41\%, with two-thirds severe) among children and adolescents.

Not surprisingly, Ethiopia consistently ranks among the world’s least peaceful nations. While recent Global Peace Index data indicate an 18.8\% improvement in the economic impact of violence in 2023, Ethiopia still ranked 144th out of 163 countries, underscoring the scale of the challenge ahead \cite{IEP2024, IEP2024PPR}. Given this context of persistent, multi-front conflict and profound human implications, there is a pressing need to move beyond descriptive accounts and adopt analytical frameworks capable of disentangling the complex, interacting drivers of violence across space and time. Robust, quantitative approaches are essential to inform policymakers and enable more effective, targeted interventions.

Despite the severity and persistence of conflict in Ethiopia, the majority of existing research remains descriptive or qualitative in nature. Seminal studies and policy reports have focused on identifying historical, political, and institutional causes of violence, often relying on narrative synthesis or high-level descriptive statistics \citep{siyum2021underlying}. While these works have advanced understanding of the structural and proximate drivers of conflict, including ethnic federalism, elite competition, and institutional weakness, they provide limited insight into the evolving spatiotemporal dynamics of violence, especially at fine geographic scales. Recent quantitative studies have begun to address the humanitarian impacts of conflict, such as disruptions to health, food security, and WASH services, using regression-based approaches and survey data \citep{abay2022access}. However, these analyses do not capture the complex spatial and temporal dependencies that characterize conflict fatalities, nor do they provide the uncertainty-aware risk estimates needed for robust early warning and resource allocation. 

To overcome these limitations, this study argues for the application of Bayesian hierarchical models. This class of models provides a principled and powerful approach for this type of analysis because Bayesian hierarchical models provide a principled and powerful approach for this type of analysis. These models treat unknown parameters as random variables, formally incorporate prior knowledge, and quantify uncertainty through posterior distributions \cite{gelman1995bayesian, ijgi13030097}. Its hierarchical structure enables the modeling of fatality counts as outcomes influenced by latent spatial dependence, temporal trends, and spatiotemporal interactions \cite{ijgi13030097}. By modeling these dependencies directly, Bayesian models overcome the unrealistic independence assumptions of simpler methods and offer a more nuanced understanding of the factors driving conflict fatalities.

The suitability of this approach is evidenced by its successful application in broader literature on crime and conflict modeling beyond Ethiopia. This literature has increasingly adopted Bayesian hierarchical spatiotemporal frameworks, which allow for rigorous quantification of spatial clustering, temporal dynamics, and covariate effects, while explicitly modeling uncertainty \citep{liu2021understanding, EgbonGayawan2025, browning2024bayesian}. Applications in other regions have demonstrated the value of these models for identifying hotspots, forecasting risk, and supporting targeted interventions. Notable examples include the use of Bayesian models to analyze crime in China \citep{liu2021understanding}, conflict fatality modeling using the point process approach in Nigeria \citep{EgbonGayawan2025}, violent conflict mapping in West and Central Africa \citep{EGBON2024100828}, public health implications of conflict-related fatality in Nigeria \citep{egbon2025spatial}, and event cascades in South Asia \citep{browning2024bayesian} using the Bayesian discrete-time Hawkes process. These studies leverage advances such as the Integrated Nested Laplace Approximation (INLA) and stochastic partial differential equation (SPDE) methods to efficiently fit complex hierarchical models to large-scale spatial data. Moreover, \citet{zens2025short} adopted a Bayesian panel model to link conflict events with displacement patterns in Somalia, demonstrating how these methods can move beyond event counting to illuminate broader human impacts.

Despite the existence of rich, geolocated event datasets like Armed Conflict Location and Event Data (ACLED) \citep{raleigh2010introducing}, there remains a conspicuous gap in the Ethiopian context: no published studies have yet applied joint Bayesian spatiotemporal models to rigorously analyze conflict fatality patterns in Ethiopia. Existing Ethiopian analyses are predominantly descriptive or cross-sectional and thus lack the methodological rigor required to uncover evolving patterns, risk factors, and uncertainty in conflict fatalities. The country's unique context, characterized by ethnic federalism, diverse topography, region-specific conflict drivers, and Ethiopia's geopolitical positioning within the highly volatile region of the Horn of Africa, remains unexplored using these advanced techniques. This omission represents a significant research opportunity to identify Ethiopian conflict hotspots, trace temporal trends, and quantify the influence of geographic and social factors.

This study addresses this gap by applying a joint Bayesian spatiotemporal modeling framework to ACLED conflict fatality data for Ethiopia (1997 -- 2024). The study integrates zero-inflated likelihoods and negative binomial dispersion to account for the data’s distributional properties, represents spatial dependence using the SPDE approximation to Mat\'ern Gaussian fields, and models temporal dynamics with autoregressive structures. Moreover, to complement the model and validate the zero-inflation component, the study fit a separate conflict fatality occurrence model. Inference is performed using INLA \citep{rue2009approximate}, which enables efficient and accurate estimation with large datasets. The outputs include high-resolution, uncertainty-aware risk maps and identification of key covariates associated with fatal outcomes intended to inform humanitarian planning, early warning, and research on the drivers of violence in Ethiopia. 

This study makes several important contributions. Methodologically, it is the first application of a dual Bayesian spatiotemporal modeling framework to the analysis of conflict fatality data in Ethiopia, offering a template for similar analyses in other complex, multi-ethnic, and polarized political environments. Practically, the resulting risk maps and quantified assessments of driver effects are valuable for evidence-based policymaking: they can inform the allocation of security and humanitarian resources, guide public health interventions, and support targeted educational and relief programs \cite{browning2024bayesian, EgbonGayawan2025}. The model’s capacity to integrate diverse data sources (such as satellite imagery, climate indicators, and mobility patterns) within a unified probabilistic framework enhances predictive power and uncertainty quantification critical for decision-making in complex environments. Ultimately, this interdisciplinary research approach bridges political science, epidemiology, and data science, translating complex patterns of violence into actionable and comprehensive insights for conflict prevention, preparedness, and response.

The remainder of this paper is structured as follows: Section \ref{sec_data} describes the data and the processing of relevant covariates. Section \ref{sec_methods} details the Bayesian hierarchical modeling framework, including the zero-inflated count model and the binary occurrence model, the specification of spatiotemporal random effects, and the inference procedure using Integrated Nested Laplace Approximations (INLA). Section \ref{sec_results} presents the findings, including model comparisons, estimated fixed effects, and the spatiotemporal patterns of fatality risk. Finally, Section \ref{sec_discussion} interprets the results in the context of Ethiopian politics and security, discusses policy implications, acknowledges limitations, suggests directions for future research, and gives a general conclusion in Section \ref{sec_conclusion}.

%%%%%%%%%%%%%%%%%%%%%%%%%
%%%%%%%%%%%%%%%%%%%%%%%%% DATA
%%%%%%%%%%%%%%%%%%%%%%%%%

\section{Data}\label{sec_data}
This study used data from the Armed Conflict Location and Event Data Project (ACLED) \cite{raleigh2010introducing}, which collects, analyzes, and disseminates detailed information on global political violence and protest events. The ACLED dataset provides information at the event level, including the date, location (longitude, latitude, and administrative units at the region, zone, and woreda levels), event and sub-event types, actor information, and reported fatalities. Events are compiled from a wide range of sources, including news agencies, official reports, and local partner networks, ensuring comprehensive coverage of conflict incidents from 1997 to the present.

For this analysis, we extracted data on violent conflict events and associated fatalities in Ethiopia from 1997 to 2024. The final dataset contained 14,271 records of conflict-related events. The unit of analysis is the individual event, with the dependent variable defined as the number of fatalities per event. Event types are classified according to ACLED’s standard user guide \cite{raleigh2015armed}. Table~\ref{desc_tab1} summarizes the distribution of conflict events, fatality counts, and the average fatalities per event.
\begin{table}[!h]
\centering
\adjustbox{max width = \textwidth}{%
\begin{tabular}{r |l c c c}
\toprule[2pt]\toprule
\textbf{Event type}& \textbf{Event (sub)} & \textbf{Event count} & \textbf{Fatality count} & \textbf{Fatality per event} \\ 
\midrule[2pt]
\multirow{3}{*}{Battles} & Armed clash & 6085 & 53266  &  8.7536  \\ 
& Non-state actor overtakes territory & 178 & 897 & 5.039  \\ 
& Government regains territory & 184 & 722 & 3.923 \\ 
\midrule

\multirow{2}{*}{Violence against civilians} & Attack & 2960 & 15392 & 5.2  \\ 
& Sexual violence & 132 & 175 & 1.3257  \\ 
%& Abduction/forced disappearance & 239 & 0 & 0  \\ %added to attack
\midrule

\multirow{1}{*}{Protests} %& Protest with intervention & 223 & 0 & 0  \\ 
%& Peaceful protest & 1794 & 0 & 0 \\ 
%& Excessive force against civilian & 500 & 1096 & 2.19  \\ 
& Excessive force against civilian & 2517 & 1096 & 0.435  \\ % all the three
\midrule

\multirow{1}{*}{Strategic development} %& Arrests & 318 & 0 & 0 \\ 
%& Non-violent transfer of territory & 62 & 0 & 0 \\ 
%& Looting/property destruction & 219 & 0 & 0 \\ 
%& Change to group/activity & 126 & 0 & 0 \\ 
%& Agreement & 142 & 0 & 0 \\ 
%& Headquarters or base established & 10 &  0 & 0  \\
%& Disrupted weapons use & 87 & 4 & 0.046 \\  
%& Other & 130 & 14 & 0.108 \\ 
& strategic development & 1094 & 18 & 0.016 \\ %All types
\midrule

\multirow{4}{*}{Remote violence} & Grenade & 122 & 154 & 1.262 \\ 
& Shelling/artillery/missile attack & 151 & 844 & 5.589 \\ 
%& Suicide bomb & 1 & 2 & 2  \\ 
& Air/drone strike & 192 & 1259  & 6.56 \\ % together with suicide bomb
& Remote explosive/landmine/IED & 86 & 391 & 4.547 \\ 
\midrule

\multirow{2}{*}{Riots} & Violent demonstration & 408 & 838 & 2.0539  \\ 
& Mob violence & 162 & 288 & 1.778 \\ 
\midrule[2pt]
\end{tabular}
}
\caption{Summary of event types and subevent types in Ethiopia, 1997 -- 2024, based on the ACLED classification \cite{raleigh2015armed}. The table reports total event counts, total fatalities, and average fatalities per event.}
\label{desc_tab1}
\end{table}

\begin{figure}[h]
\centering
\includegraphics[width=1\linewidth]{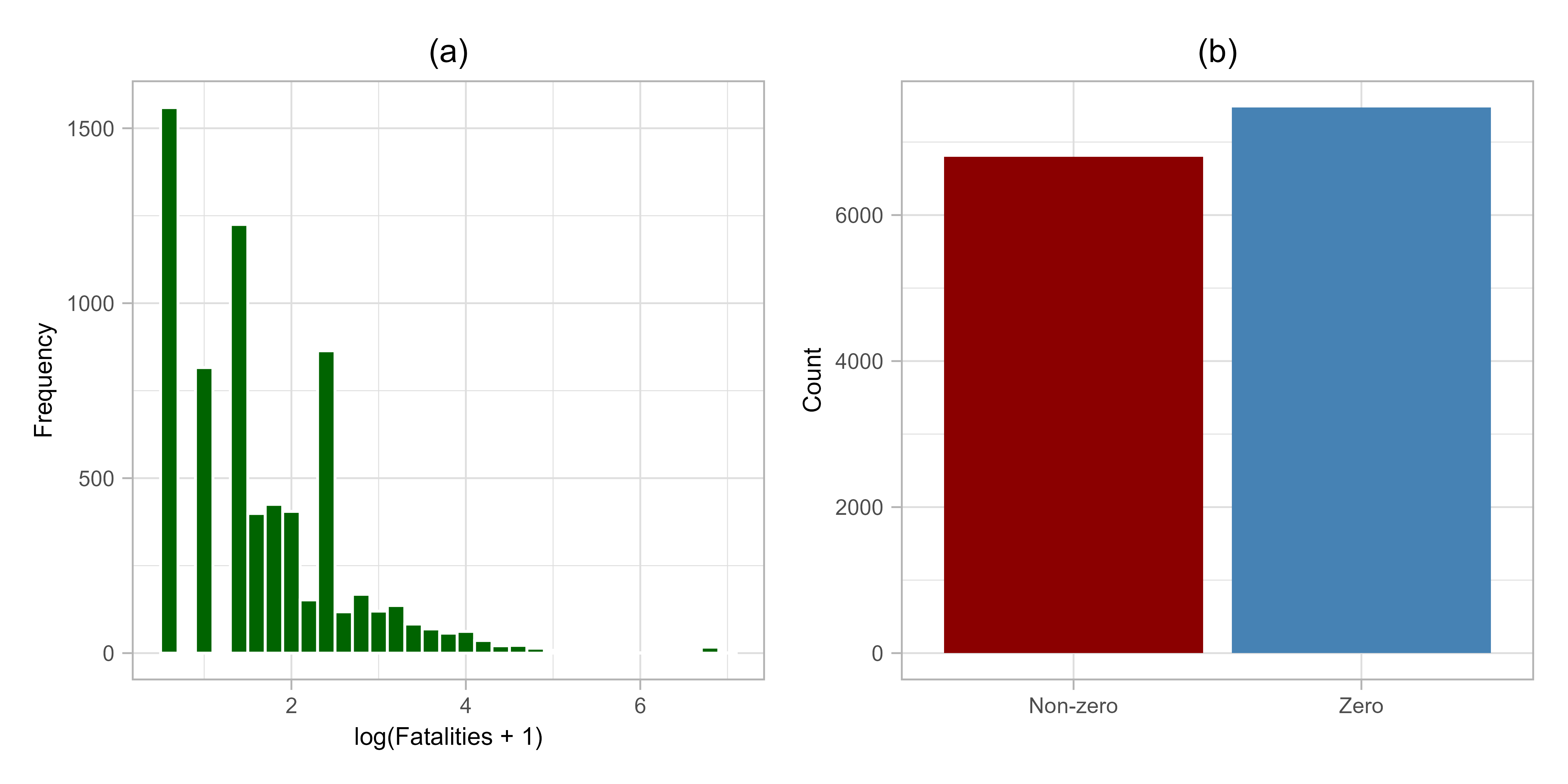}
\caption{Distribution of fatalities per event: histogram of positive fatalities on a log scale (a); proportion of events with zero versus non-zero fatalities (b).}
\label{desc_figfreq}
\end{figure}

Figure \ref{desc_figfreq} illustrates the distribution of conflict-related fatalities in Ethiopia. Panel (a) shows the frequency distribution of positive fatalities (log-transformed) to account for the highly skewed nature of data, where most events involve only a few deaths, while a small number of events involve extremely high fatalities. Panel (b) highlights the prevalence of zero fatalities, which account for more than half of all observations. Together, these plots emphasize two features of data: a high proportion of zero-fatality events and a long-tailed distribution of positive fatalities.

Using Vuong’s procedure \cite{vuong1989likelihood}, we tested for zero inflation relative to standard Poisson and negative binomial regression models with event type, season, and distances from the nearest city and border as predictors. The tests indicated significant zero inflation for both count distributions. Exploratory summaries further show that 52.4\%  of events recorded zero fatalities, while positive fatality counts exhibited extreme overdispersion (mean = 5.28, variance = 1338.63, maximum = 1172). These features motivate the use of a zero-inflated and overdispersed spatiotemporal count model. 

\begin{figure}[!h]
\centering
\includegraphics[width=1\linewidth]{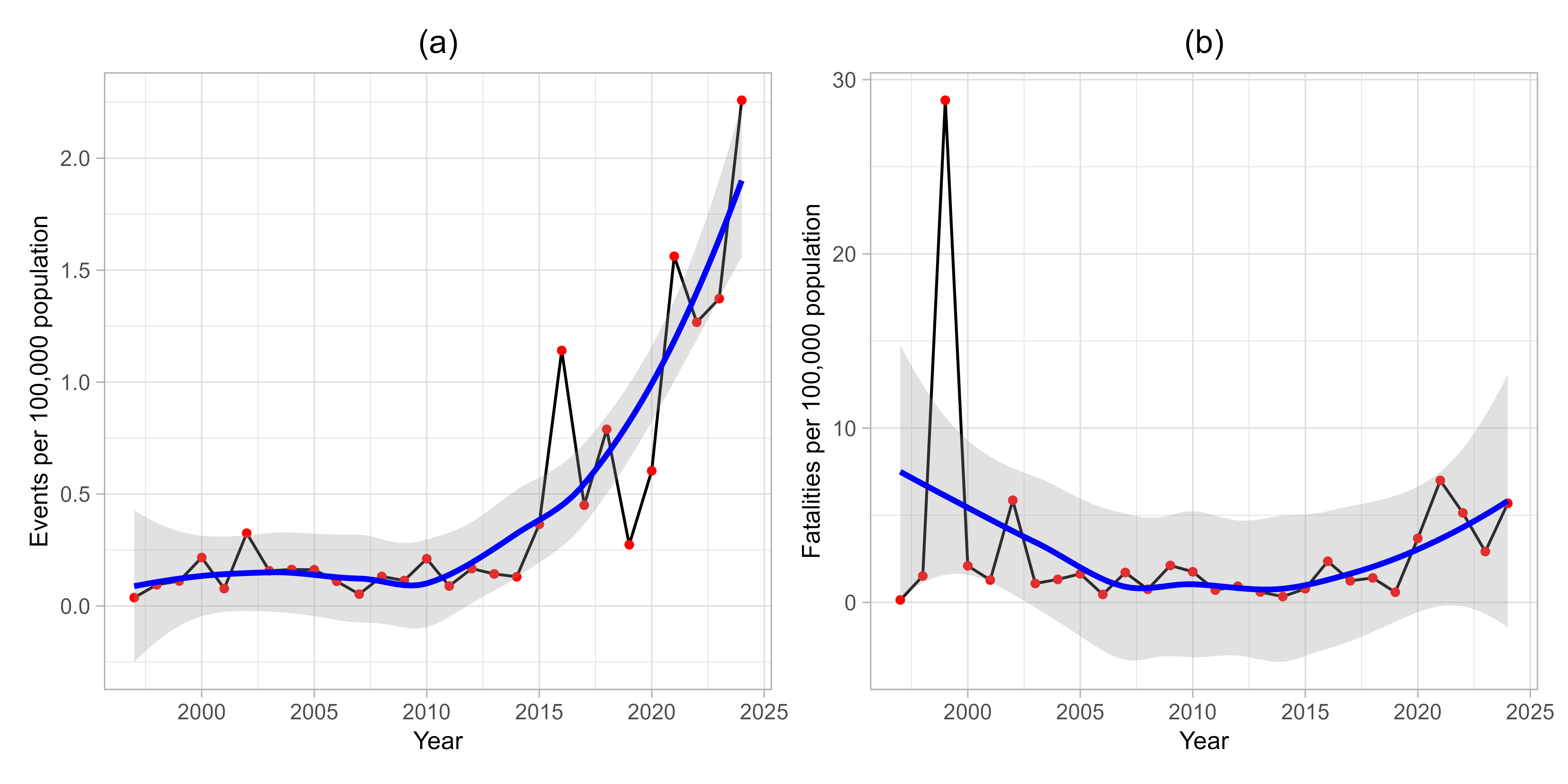}
\caption{Rates of violent conflict events (a) and fatalities (b) per 100,000 population across years, Ethiopia 1997 -- 2024. The figures were generated by the authors in \texttt{R} version 4.5.1 using the ggplot2 (version 4.0.0) package.}
\label{desc_figpop}
\end{figure}

\begin{figure}[!h]
\centering
\includegraphics[width=1\linewidth]{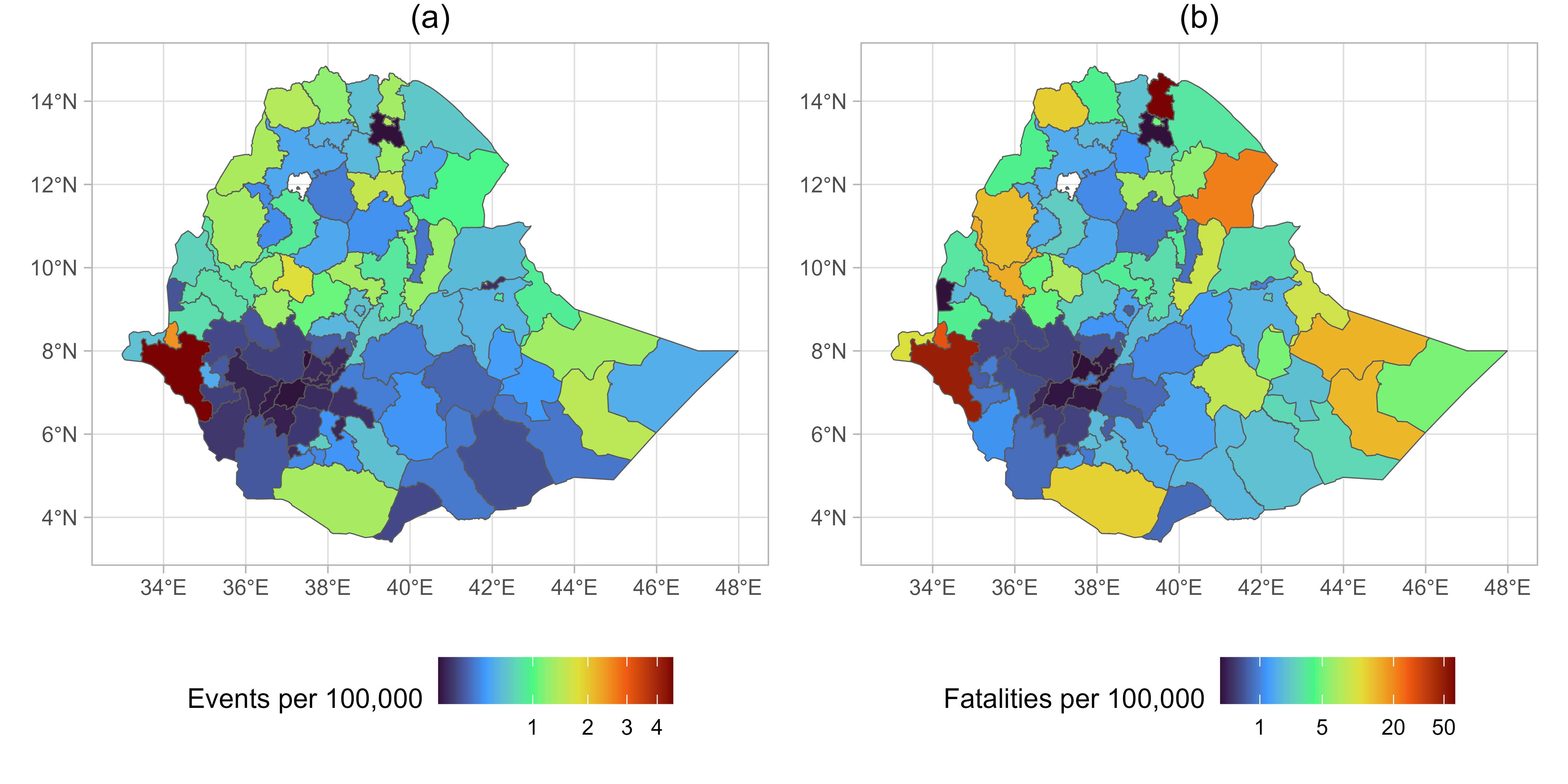} %desc_spat_rate1
\caption{Rates of violent conflict events (a) and fatalities (b) per 100,000 population across zonal administrative regions, Ethiopia 1997 -- 2024. The figures were generated by the authors in \texttt{R} version 4.5.1 using the ggplot2 (version 4.0.0) package.}
\label{desc_figspatpop}
\end{figure}

Temporal and spatial descriptive statistics are presented in Figures~\ref{desc_figpop} and \ref{desc_figspatpop}. Figure~\ref{desc_figpop} shows the annual rates of conflict events and fatalities (per 100,000 population) from 1997 to 2024. The temporal pattern is multimodal, with an initial peak during the Ethio -- Eritrean War (1998 -- 2000), followed by a period of relative stability, and then a sharp increase beginning around 2015/16, peaking between 2020 and 2022. The parallel rise in both events and fatalities in this later period reflects a marked escalation in conflict intensity.
Figure~\ref{desc_figspatpop} presents the spatial distribution of conflict events and fatalities per 100,000 population across zonal administrative regions. Events are concentrated in the north and west, particularly along Ethiopia’s western border. Fatalities show an even sharper concentration, with pronounced hotspots in the north, west, and east. These sharper concentrations of fatalities of conflicts are along the country's international border with the northern (Eritrea), western (Sudan), and eastern (Somalia). This suggests that while conflict occurs in many areas, its lethality is highly localized.

Finally, the study incorporated both categorical and continuous covariates into the hierarchical spatiotemporal model of fatality counts and occurrences as fixed effects. Categorical covariates included the type of conflict event (Table~\ref{desc_tab1}), primary actors involved in the violent conflicts, such as state forces, rebel groups, communal identities, militia groups, rioters, protesters, and civilians; and the season (autumn, winter, spring, and summer). For more detailed explanations for event types and actors, see the ACLED codebook \cite{raleigh2015armed}. Continuous predictors were the distances from each event to the nearest city and to the international border (in kilometers), modeled as smooth nonlinear functions using penalized splines (RW(2)). Temporal correlation was captured via an autoregressive (AR(1)) year effect, while spatial dependence was modeled with a Mat\'ern field evolving over time. Population data from WorldPop \citep{Worldpop2025} were incorporated as an offset to standardize fatality counts across regions of differing population sizes.

%%%%%%%%%%%%%%%%%%%%%%%%%
%%%%%%%%%%%%%%%%%%%%%%%%% METHODS
%%%%%%%%%%%%%%%%%%%%%%%%%

\section{Method}\label{sec_methods}
%\subsection{Model overview and notation}
Let $Y_{it} \in \mathbb{N}_0$ denote the fatality count associated with a violent conflict event $i = 1, \dots, n$, occurring at spatial location $s_i \in \mathcal{D} \subset \mathbb{R}^2$ and time $t \in \mathcal{T} \subset \mathbb{R}^+$, with observed value $y_{it}$. We model $Y_{it}$ using a count distribution with a mean parameter $\mu_{it} > 0$, augmented by a mechanism for generating structural zeros. Specifically, our modeling framework consists of two linked components: a count component that models the distribution of positive counts (as well as sampling zeros) and a zero-inflation component that models the probability of a structural zero. Both components incorporate fixed-effect covariates, spatial and temporal random effects, and a spatiotemporal interaction term. The count mean $\mu_{it}$ is linked to its linear predictor via a log link function, while the zero-inflation probability $\psi_{it}$ is linked via a logit link. An offset term is included where appropriate to account for population at risk or exposure for the count component.

\subsection{Zero-inflated count models}
Count data in many applications contain more zero observations than standard distributions, such as the Poisson or negative binomial, can accommodate. This phenomenon, known as zero-inflation, can lead to poor model fit and biased inference if ignored, as standard models underestimate the probability of zeros. To address this, we employ zero-inflated (ZI) models, which explicitly account for zeros arising from two distinct sources \citep{lambert1992zero}. Our fatality count data exhibit a substantial proportion of zeros, exceeding the predictions of conventional count models, making the ZI framework appropriate. In a ZI model, an observation can be a structural zero, which occurs with a certain probability independently of the count process, or a sampling zero, generated by the underlying count distribution. 

Formally, the ZI model is specified as a mixture distribution given by
\begin{equation}\label{zi_general}
\pi(Y_{it}=y_{it}\mid\mu_{it},\psi_{it}) =
\begin{cases}
	\psi_{it} + (1-\psi_{it})\,\pi(0\mid\mu_{it}), & y_{it} = 0,\\
	(1-\psi_{it})\,\pi(y_{it}\mid\mu_{it}), & y_{it} > 0,
\end{cases}
\end{equation}
where $\psi_{it}$ is the probability that event $i$ at time $t$ is a structural zero, and $\pi(\cdot\mid\mu_{it})$ is an untruncated count distribution with parameter $\mu_{it}$. For our data, structural zeros represent events not expected to produce fatalities (e.g., peaceful demonstrations), captured by the first component of Eq.~\eqref{zi_general}. The second component accounts for sampling zeros, which occur by chance (e.g., a violent event with no recorded fatalities). We model the expected fatality count parameter $\mu_{it}$ and probability
of a structural zero $\psi_{it}$ (represents the prevalence parameter indicative of the proportion of excessive zeros, also referred to as the mixture probability) using a log and logit link function, respectively, 
\begin{equation}\label{eq_ZI}
	\log(\mu_{it}) = \eta^c_{it} + \log(E_{it}), ~~ \text{logit}(\psi_{it}) = \eta^{\psi}_{it},
\end{equation}
where $\eta^c_{it}$ and $\eta^\psi_{it}$ are corresponding linear predictors for each component. $E_{it}$ is an offset representing the population at risk in a spatial and spatiotemporal model that satisfies $\log(\mu_{it}/E_{it}) = \eta^c_{it}$ \cite{EGBON2024100828}. The choice of count distribution $\pi(\cdot)$ is important. While the zero-inflated Poisson (ZIP) model is common, it assumes equality of the conditional mean and variance, an assumption often violated in real-world data by overdispersion. Overdispersion can arise from population heterogeneity, unobserved covariates, or outliers \citep{dean2014overdispersion, payne2017approaches, feng2021comparison}. The zero-inflated negative binomial (ZINB) model relaxes this assumption and is therefore often preferred for overdispersed data. The selection of likelihood for the count process involved choosing the ZINB distribution over the ZIP was conducted using the Deviance Information Criterion (DIC) \cite{spiegelhalter2002bayesian}, which indicated that the ZINB (DIC = 56245.85) model outperformed the ZIP (DIC = 58951.03), as evidenced by a lower DIC value.

\subsection{Occurrence model for binary outcomes}\label{sec_occ_model}
To complement the ZINB model and validate the zero-inflation component, we fit a separate fatality occurrence and non-occurrence model \citep{lambert1992zero, egbon2025spatial} analyzing the probability of an event resulting in at least one fatality. Let 
\begin{equation}
	O_{it} = \mathbbm{1}(Y_{it} > 0),
\end{equation}
be the binary indicator, which takes the value 1 if an event $i$ at time $t$ had one or more fatalities and 0 otherwise, modeled using a Bernoulli distribution as
\begin{equation}\label{eq_occ_mod}
\begin{split}
	O_{it} &\sim \mathrm{Bernoulli}(p_{it}),\\
	\text{logit}(p_{it}) &= \log\left(\frac{p_{it}}{1-p_{it}}\right) =  \eta^{(o)}_{it},
\end{split}
\end{equation}
where $p_{it}$ is the probability of a fatality occurrence. This model decouples the occurrence process from the magnitude of fatalities, allowing robust assessment of spatiotemporal risk patterns and serving as a validation tool for the zero-inflated model.

\subsection{Structure of the linear predictors and prior specifications} \label{sec_linear_pred}
The linear predictors for the count mean ($\eta^c_{it}$), zero-inflation probability ($\eta^{\psi}_{it}$), and binary occurrence probability ($\eta^o_{it}$) capture structured spatiotemporal variation not explained by covariates. For the count component, the linear predictor is given by
\begin{equation}\label{lin_pred}
\begin{split}
	\eta^c_{it} &= \mathbf{z}_{it}^{\top}\boldsymbol{\beta} + \sum_{j=1}^{2} f_j\left(d_i^{(j)}\right) + v_t + u(s_i) + w(s_i,t),
\end{split}
\end{equation}
where $\mathbf{z}_{it} = (1, z_{1it}, z_{2it} \cdots z_{pit})^{\top}$ is a vector of $p$ covariates with associated coefficients $\boldsymbol{\beta}$, $f_j(\cdot)$ are smooth functions of distance-related variables from the international border, $v_t$ is a temporal random effect, $u(s_i)$ is a static spatial effect, and $w(s_i,t)$ is a spatiotemporal interaction. The temporal effect $v_t$ is modeled as autoregressive of order one (AR(1)), the spatial effect $u(s)$ as a Gaussian random field (GRF) with Mat\'ern covariance using the SPDE approach \citet{lindgren2011explicit}, and the interaction $w(s,t)$ as a dynamic GRF evolving through AR(1) in time and Mat\'ern spatial dependence. Smooth terms $f_j(\cdot)$ are represented either as random walks of order two (RW2). 

The appendix contains the complete theoretical formulation of prior specifications for continuous spatial ($u(s_i)$) and spatiotemporal ($w(s_i,t)$) models, as well as details on the nonlinear functions in time ($v_t$) and distance ($f_j(\cdot)$). A similar formulation can be applied to the zero-inflation and binary components ($\eta^\psi_{it}$ and $\eta^o_{it}$), with potentially different coefficient sets. Alternative model specifications explored in Section~\ref{sec_model_variants} include reduced versions of Eq. \eqref{lin_pred} that omit some of the latent terms.

A fully Bayesian model requires prior distributions for all unknown parameters, including fixed effects and the hyperparameters governing the random effects. Our choice of priors aims to balance regularity with computational stability, favoring weakly informative priors that constrain parameters to plausible ranges without overly influencing the posterior estimates \citep{lindgren2015bayesian}. We follow the penalized complexity (PC) prior framework \citep{simpson2017penalising}, and adopt the calibration strategy of \citet{egbon2025spatial} for the SPDE model hyperparameters of the spatial range $\rho$ and the marginal standard deviation $\sigma_\omega$. Specifically, we define the PC priors as
\begin{equation*}
	\pi(\rho < \rho_0) = p_\rho, \qquad \pi(\sigma > \sigma_0) = p_\sigma,
\end{equation*}
where $\rho_0$ is a lower bound for the spatial range and $\sigma_0$ is an upper bound for the marginal standard deviation. To motivate $\rho_0$ in the Ethiopian context, note that the largest administrative zone covers approximately $46{,}417$ km$^2$. Approximating this zone as a circle gives a radius of 121.6 km. Since $1^\circ$ latitude corresponds to roughly $111$ km, this equates to about $1.1^\circ$. We therefore set $\pi(\rho < 200) = 0.9$, reflecting the belief that spatial dependence is unlikely to decay within distances smaller than a typical administrative zone. For the marginal standard deviation, we set $\pi(\sigma > 3) = 0.01$, which constrains the variability to realistic levels while maintaining sufficient flexibility. PC priors \citep{simpson2017penalising} are used for all the AR(1) correlation parameters. For the fixed-effect coefficients ($\boldsymbol{\beta}$), we used the default \texttt{R-INLA} Gaussian priors. Finally, when a zero-inflated negative binomial (ZINB) likelihood is used for the count response, the dispersion parameter $\theta$ (with $\alpha = 1/\theta$) is assigned the default \texttt{R-INLA} prior.  

\subsection{Bayesian inference}\label{sec_inference}
Inference for the zero-inflated spatiotemporal count model is performed within a Bayesian framework. The complex hierarchical structure, including the spatial and spatiotemporal Gaussian random fields (GRFs), places this model within the class of latent Gaussian models (LGM) \cite{rue2009approximate}, for which the Integrated Nested Laplace Approximation (INLA) provides a computationally efficient and accurate alternative to traditional Markov Chain Monte Carlo (MCMC) methods. All models were estimated using the \texttt{R-INLA} package \citep{rue2009approximate}, which is specifically designed for this task. Let $\mathbf{x} = (\boldsymbol{\beta}, \mathbf{f}, \mathbf{v}, \mathbf{u}, \mathbf{w})$ denote the latent field comprising fixed effects, smooth functions, and random effects, and $\boldsymbol{\theta}$ be the hyperparameters governing precisions and dependence structures. The joint posterior distribution factors as
\begin{equation}
\pi(\mathbf{x}, \boldsymbol{\theta} \mid \mathbf{y}) \propto \prod_{i=1}^{n}\prod_{t=1}^{T}\pi(y_{it} \mid \mathbf{x}, \boldsymbol{\theta}) \times \pi(\mathbf{x} \mid \boldsymbol{\theta}) \times \pi(\boldsymbol{\theta})
\end{equation}
where $\pi(y_{it} \mid \mathbf{x}, \boldsymbol{\theta})$ is the likelihood given in Eq. \eqref{zi_general} for $i = 1, \cdots, n$, and $t = 1997,\cdots, 2024$,\ $\pi(\mathbf{x} \mid \boldsymbol{\theta})$ represents the joint prior on the latent field, and $\pi(\boldsymbol{\theta})$ are the hyperpriors. The marginal posterior distribution of the latent field $\mathbf{x}$ is obtained by integrating out $\boldsymbol{\theta}$ from the joint posterior distribution, while the marginal posterior distribution of $\boldsymbol{\theta}$ is obtained by integrating out $\mathbf{x}$ from the joint distribution. The primary goal is to compute the marginal posterior distributions for the latent field, $\pi(x_j \mid \mathbf{y})$, and the hyperparameters, $\pi(\theta_k \mid \mathbf{y})$ (for details, see \citep{rue2009approximate,lindgren2015bayesian}). 

\subsection{Model variants and selection}\label{sec_model_variants}
Given the complexity of the latent process, we fit a suite of models to identify the specification that best captures the spatiotemporal dynamics of conflict fatalities in Ethiopia. The models vary along the inclusion of specific random effects components. The structure of the main models considered is summarized in Table \ref{tab_model_specs}. For all models, the linear predictor for the zero-inflation component mirrors the structure of the count component. We also consider the occurrence model (Eq.~\eqref{eq_occ_mod}) for validation of the ZI component.
\begin{table}[!h]
\centering
\adjustbox{max width = \textwidth}{%
\begin{tabular}{r| c l c }
\toprule[2pt]\toprule
& \textbf{Model identifier} & \textbf{Linear predictor} & \textbf{Likelihood} \\ 
\midrule[2pt]
\multirow{3}{*}{Count} & CM1 & $\eta^c_{it} = \mathbf{z}_{it}^{\top}\boldsymbol{\beta} + f_1\left(d^{(1)}_i\right) + f_2\left(d^{(2)}_i\right) + v_t + u(s_i)$ &   \\ 
& CM2 & $\eta^c_{it} = \mathbf{z}_{it}^{\top}\boldsymbol{\beta} + f_1\left(d^{(1)}_i\right) + f_2\left(d^{(2)}_i\right) + w(s_i,t)$ & ZINB \\ 
& CM3 & $\eta^c_{it} = \mathbf{z}_{it}^{\top}\boldsymbol{\beta} + f_1\left(d^{(1)}_i\right) + f_2\left(d^{(2)}_i\right) + v_t + w(s_i,t)$ &  \\ 
\midrule
\multirow{3}{*}{Occurrence} & OM1 & $\eta^o_{it} = \mathbf{z}_{it}^{\top}\boldsymbol{\beta} + f_1\left(d^{(1)}_i\right) + f_2\left(d^{(2)}_i\right) + v_t + u(s_i)$ &  \\ 
& OM2 & $\eta^o_{it} = \mathbf{z}_{it}^{\top}\boldsymbol{\beta} + f_1\left(d^{(1)}_i\right) + f_2\left(d^{(2)}_i\right) + w(s_i,t)$ & Bernoulli \\ 
& OM3 & $\eta^o_{it} = \mathbf{z}_{it}^{\top}\boldsymbol{\beta} + f_1\left(d^{(1)}_i\right) + f_2\left(d^{(2)}_i\right) + v_t + w(s_i,t)$ &  \\
\midrule[2pt]
\end{tabular}
}
\caption{Specification of key Bayesian hierarchical models fitted to the conflict fatality data. The notation follows Equations \eqref{eq_ZI} and \eqref{lin_pred}.}
\label{tab_model_specs}
\end{table}

To evaluate model fit and predictive performance, we used several Bayesian and classical diagnostics. Model selection and comparison were performed using Deviance Information Criterion (DIC) \cite{spiegelhalter2002bayesian}, Watanabe-Akaike information criterion (WAIC) \citep{watanabe2013widely}, and the negative mean logarithm of conditional predictive ordinate (LCPO) \citep{roos2011sensitivity}. Here, we LCPO is calculated using $LCPO =\frac{1}{nT}\sum_{t=1}^{T}\sum_{i=1}^{n_t}\log(CPO_{it})$ where $CPO_{it}$ are CPO values for the observation $i$ at time $t$. Lower values of DIC, WAIC, and LCPO indicate better fit and predictive performance. The best-performing model was used for final inference and interpretation.

%%%%%%%%%%%%%%%%%%%%%%%%%%%%
%%%%%%%%%%%%%%%%%%%%%%%%%%%%   %Result
%%%%%%%%%%%%%%%%%%%%%%%%%%%%

\section{Result}\label{sec_results}
This section presents the findings from our analysis of violent conflict fatalities in Ethiopia from 1997 to 2024. To disentangle the complex processes driving lethality, we fitted two complementary Bayesian hierarchical spatiotemporal models using the INLA-SPDE approach. Both models accounted for latent dependencies through a structured Gaussian random field (spatial effect), a first-order autoregressive process (temporal effect), and a smooth function along distances from the nearest border. They also shared a common set of fixed effects such as season, event type, and distance to the nearest city. The models were distinguished by their response distributions to answer distinct research questions. First, a ZINB model analyzed the count of fatalities per event to identify factors influencing the scale of lethality. Second, a binomial model analyzed the binary occurrence of any fatalities to identify factors influencing the risk or probability that of a violent conflict event becomes lethal. This dual-model framework provides a more nuanced understanding than a single model alone, distinguishing the drivers of whether an event is lethal from the drivers of how lethal it is.

The results are structured in the following manner. To ensure that our method works, we first compare the model performances using the DIC, WAIC, and LCPO (Section \ref{mod_comparison}). Thereafter, in Section \ref{mod_summary}, we show and explain the posterior estimates for the fixed and random effects, going into detail about the main causes and spatiotemporal patterns of conflict lethality in Ethiopia.

\subsection{Model comparison} \label{mod_comparison}
To select the optimal model specification, we evaluated competing models using the DIC, WAIC, and the LCPO. These criteria penalize model complexity to avoid overfitting, with lower values indicating a superior balance of fit and predictive accuracy.
\begin{table}[!h]
\centering
\adjustbox{max width = \textwidth}{%
\begin{tabular}{r| c c c c c}
\toprule[2pt]\toprule
& \textbf{Model identifier} & \textbf{Likelihood} & \textbf{DIC} & \textbf{WAIC} & \textbf{LCPO} \\ 
\midrule[2pt]
\multirow{3}{*}{Count} & CM1 & ZINB & 55054.58 & 56234.15 & 1.957 \\ 
& CM2 & ZINB & 54393.87 & 55076.35 & 1.180 \\ 
& CM3* & ZINB & 53692.15 & 53955.25 & 1.047 \\ 
\midrule[1.25pt]
\multirow{3}{*}{Occurrence} & OM1 & Bernoulli & 15031.90 & 15043.06 & 0.542 \\ 
& OM2 & Bernoulli & 14286.37 & 14268.44 & 0.496 \\ 
& OM3* & Bernoulli & 14248.74 & 14224.50 & 0.429 \\
\midrule[2pt]
\end{tabular}}
\caption{Model performance and comparison of models specified in Table \ref{tab_model_specs}. A preferable model is characterized by lower DIC, WAIC, and LCPO values.}
\label{tab_mod_comp}
\end{table}

As shown in Table \ref{tab_mod_comp}, for modeling the count of fatalities, CM3 (DIC = 53692.15, WAIC = 55034.59, LCPO = 1.047) was the preferred model based on its lower values of DIC, WAIC, and LCPO compared to CM1 and CM2. Similarly, for modeling the occurrence of fatalities, OM3 demonstrated the best performance, yielding the lowest values across all three criteria (DIC = 14248.74, WAIC = 14224.50, LCPO = 0.429). Consequently, models CM3 and OM3 were selected for all subsequent inference and interpretation.

\subsection{Model summary} \label{mod_summary}
This section presents the posterior estimates for the fixed effects, random effects, and hyperparameters from our final models. The fatality count data were modeled using a ZINB distribution (CM3), while the occurrence of any fatality was modeled with a binomial distribution (OM3), as selected in Table \ref{tab_mod_comp}.

\subsubsection{Effect of fixed effects on fatality}
\subsubsection*{a. Event types}
\begin{figure}[!h]
\centering
\includegraphics[width=1\linewidth]{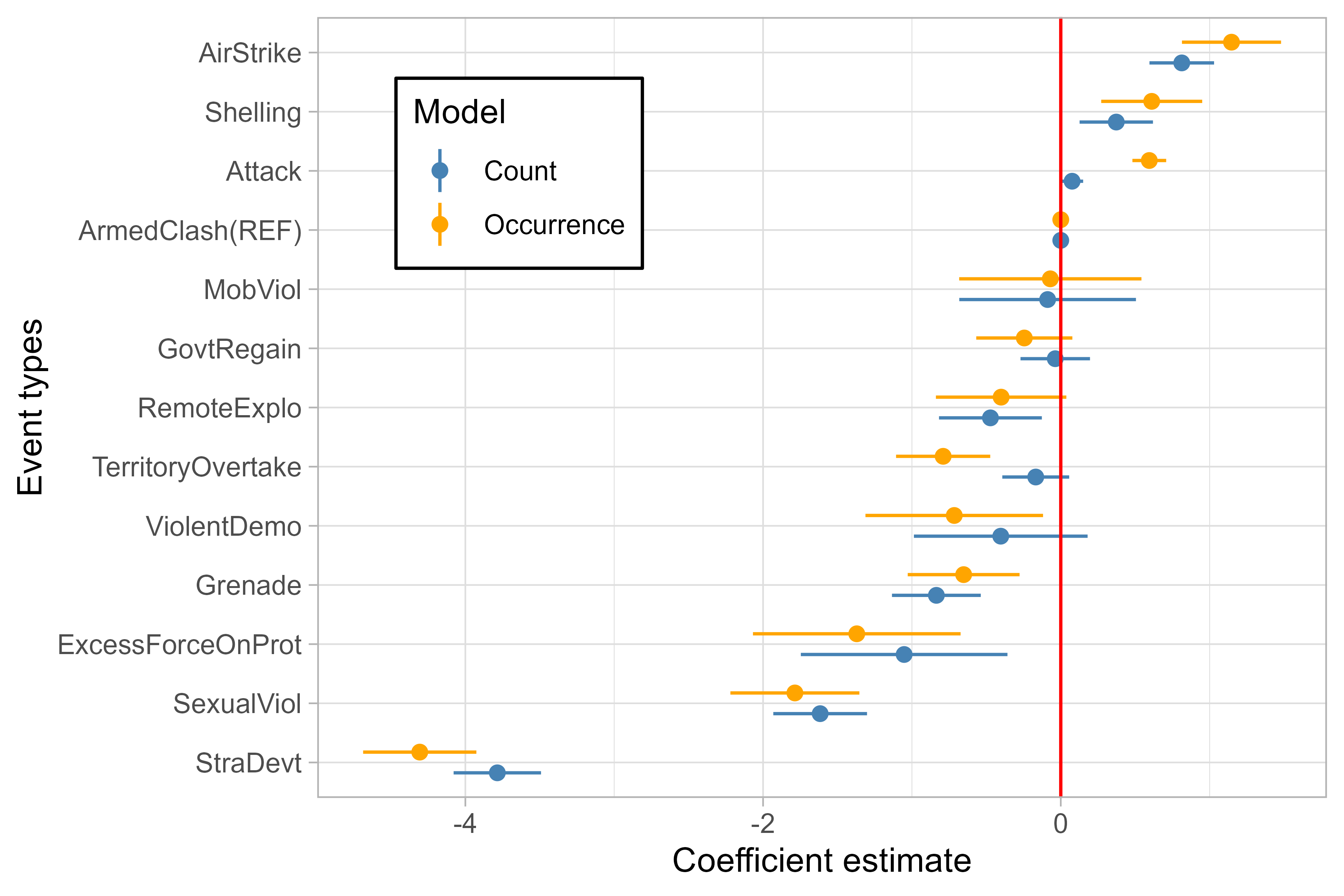}
\caption{Fixed effect coefficients for different event types relative to armed clash ("ArmedClash") (reference category) from two models: a ZINB model (blue) for conflict fatality count and a binomial model (orange) for fatality presence/absence. Points represent posterior means, and bars indicate 95\% credible intervals.}
\label{event_type}
\end{figure}

The type of violent event emerged as one of the most substantial predictors of both the occurrence and intensity of fatalities. Figure \ref{event_type} presents the posterior mean and 95\% credible intervals (CIs) for the influence of event type, representing the log-odds of fatality occurrence (binomial model) and the log of the expected count (ZINB model), using "Armed clash" as the reference category. A detailed definition and description of these event types can be found in the ACLED codebook \cite{raleigh2015armed}.

In the occurrence model, event types demonstrated distinct effects on the probability of fatality occurrence. Positive effects were observed for \textit{air strikes, shelling/artillery/missile attacks}, and \textit{attacks}, indicating these events significantly increase the likelihood of fatalities compared to \textit{armed clashes}. Negative effects were found for \textit{strategic developments, sexual violence, excessive force against protesters, grenades, violent demonstrations}, and \textit{non-state actors overtaking territory}, suggesting these events reduce fatality probability. 
This result corroborates earlier evidence and expert assessments that associate the use of modern or sophisticated warfare with higher fatality rates, highlighting the devastating impacts of explosive weapons \citep{bagshaw20232022, gebregziabher2025civilian}. 
Non-significant effects were observed for \textit{remote explosions, mob violence}, and \textit{government regaining territory}, as their credible intervals included zero. 

Similarly, from the count model, event types showed different patterns for fatality intensity. \textit{Air strikes, shelling/artillery/missile attacks}, and \textit{attacks} showed strong, significant positive effects, indicating these event types are significantly more likely to result in fatalities and produce higher fatality counts when occurring. Negative effects were observed for \textit{strategic developments, sexual violence, excessive force against protesters, grenades}, and \textit{remote explosions}, suggesting reduced fatality counts. Non-significant effects were found for \textit{mob violence, government regaining territory, violent demonstrations}, and \textit{non-state actors overtaking territory} from the reference category.

The analysis reveals three patterns relative to armed clashes: air strikes, shelling, and attacks increase both fatality occurrence and intensity, while sexual violence, strategic developments, excessive force against protesters, and grenade attacks decrease both dimensions compared to the reference. Remote explosions reduce only fatality counts, and territory overtakes reduce only occurrence probability. This stratification demonstrates distinct mechanistic pathways relative to armed clashes; some events increase fatality, others decrease it, and a few show dimension-specific effects.

\subsubsection*{b. Actors}
\begin{figure}[!h]
\centering
\includegraphics[width=1\linewidth]{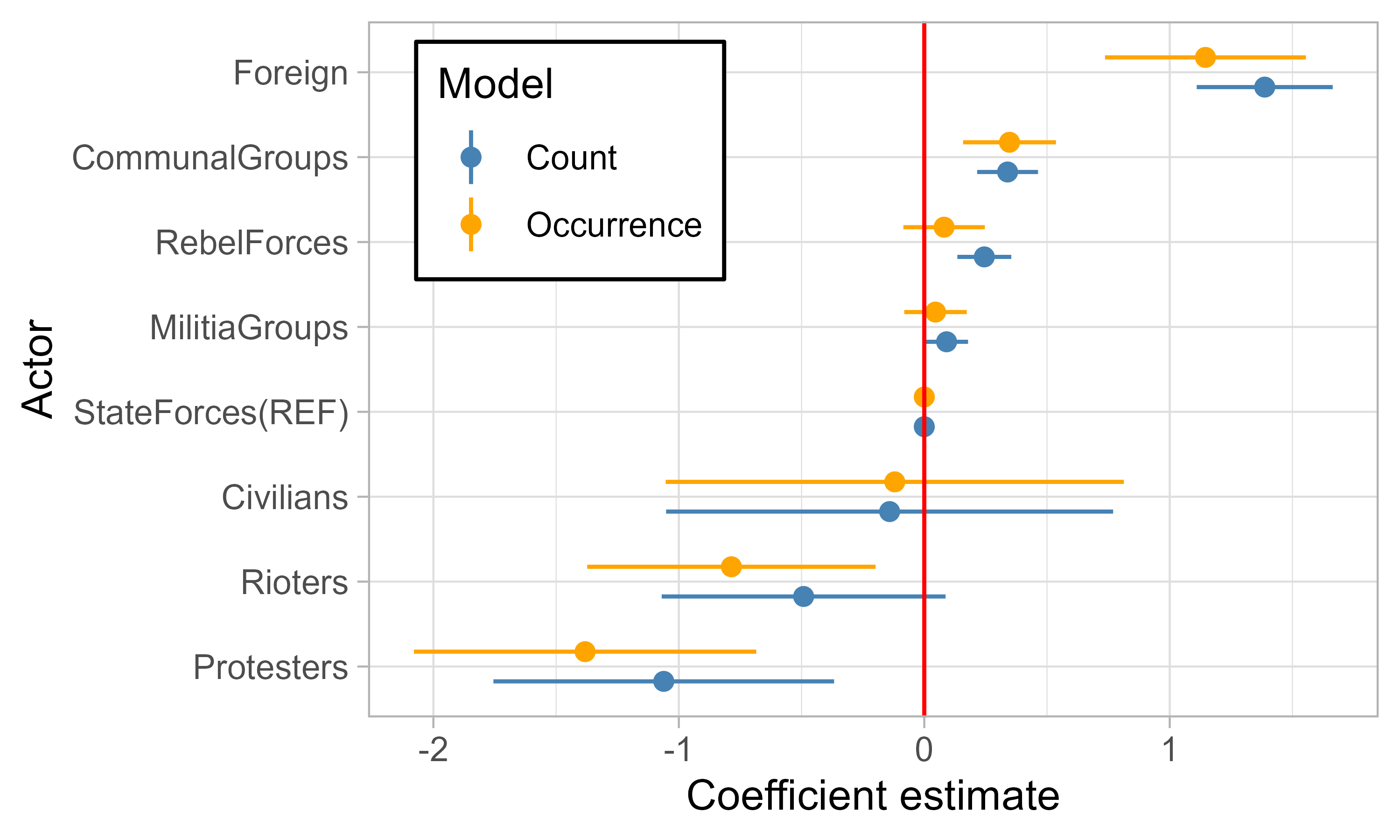}
\caption{Fixed effect coefficients for different actors relative to state forces ("StateForces") (reference category) from two models: a ZINB model (blue) for conflict fatality count and a binomial model (orange) for fatality presence/absence. Points represent posterior means, and bars indicate 95\% credible intervals.}
\label{actors}
\end{figure}

ACLED classifies actors into state forces, rebels, militias, communal groups, demonstrators, civilians, and external forces \cite{raleigh2015armed}. Organized armed actors such as governments, rebels, and militias typically pursue broader political objectives, while rioters and protesters are less structured and often act spontaneously. Civilians are usually recorded as victims rather than active participants. The results shown in Figure \ref{actors} summarize the estimated effects of different actor types on conflict fatalities and occurrences relative to state forces (reference).

The result revealed that the \textit{foreign groups} are ranked as the top actors with the highest odds of fatality occurrence by 215\% (odd ratio = 3.15) compared to state forces, as shown in Figure \ref{actors}. This indicates that violent events that involve foreign groups are more likely to lead to fatality occurrence compared with the reference. Empirical research supports these findings; for example, analysis of external interventions demonstrates higher conflict intensity and greater lethality when foreign or external actors intervene directly or indirectly in internal conflicts \citep{schmiedl2021conditions, demena2024meta}. These studies describe how such involvement tends to change the scale of the conflict, the type of weapons used, and the overall fatality. \textit{Communal/identity groups} also significantly increase the probability of fatalities (41\% higher odds). Conversely, \textit{protesters} show the most substantial protective effect, with 75\% lower odds of fatalities occurring, followed by \textit{rioters} with 54\% lower odds. Notably, \textit{rebel forces} and \textit{militia groups}, while influential in the count model, show no statistically significant effect on the basic occurrence of fatalities, suggesting their impact manifests differently in conflict dynamics.

Similarly for the count model, which examines the number of fatalities when they occur, a more comprehensive hierarchy of lethality emerges. \textit{Foreign actors} again show the strongest effect, associated with four times higher fatality counts compared to state forces. \textit{Communal groups} and \textit{rebel forces} both significantly increase fatality intensity with 40\% and 28\% higher counts, respectively, while \textit{militia groups} show a more modest but still significant effect with a 10\% increase. The protective effects for \textit{protesters} and \textit{rioters} remain evident, with 65\% and 39\% fewer fatalities, respectively; this is predominantly because of the modalities and means used by  protest movements and riots are relatively peaceful and less destructive.  

\subsubsection*{c. Season}
From June to September, Ethiopia experiences its main rainy season (kiremt), while October to February is generally dry (bega). To align with standard analytical frameworks, we classified the seasons as spring (March -- May, part of the short rains, belg, and dry season), summer (June -- August, peak rainy season), autumn (September -- November, end of rains and start of dry season), and winter (December -- February, dry season).

The seasonal variation shows a consistent peak in summer across both model components, as shown in Table \ref{res_seasonal_effect}. Summer is associated with a significantly higher likelihood of fatality occurrence and a substantial increase in the expected number of fatalities compared to the reference season (spring).
\begin{table}[!h]
\centering
\adjustbox{max width = \textwidth}{%
\begin{tabular}{lcccccccc}
\midrule[2pt]
& \multicolumn{3}{c}{Count} &&& \multicolumn{3}{c}{Occurrence} \\
\cline{2-4} \cline{7-9}
& Mean & 2.5\% & 97.5\% &&& Mean & 2.5\% & 97.5\% \\
\midrule[1pt]
\textit{Spring} & \multicolumn{3}{c}{Reference} &&& \multicolumn{3}{c}{Reference} \\
Autumn & 0.039 & -0.049 & 0.127 &&& -0.046 & -0.171 & 0.080 \\
Winter & -0.028 & -0.116 & 0.061 &&& -0.061 & -0.188 & 0.066 \\
Summer & 0.228 & 0.141 & 0.316 &&& 0.170 & 0.043 & 0.297 \\
\midrule[1.25pt]
\end{tabular}}
\caption{Posterior means and 95\% credible intervals for seasonal effects on  and fatality counts (Count) and fatality occurrence (Occurrence). Spring is used as the reference.}
\label{res_seasonal_effect}
\end{table}
The effects for Autumn and Winter are both small and non-significant, as their 95\% credible intervals include zero. This indicates that the probability and intensity of fatal events during these seasons are not meaningfully different from those in spring. Together, these results indicate that the main rainy season (summer) is the period of peak conflict risk in Ethiopia, characterized by both a higher probability of fatal events and more severe outcomes when they occur. 
The increasing occurrences and intensity observed during this season can be attributed to environmental conditions that favor insurgent activity. The rainy weather provides cover and mobility advantages for armed groups, allowing them to operate more freely and reduce their exposure to aerial surveillance, drone strikes, and mechanized state forces. It also facilitates the movement and smuggling of weapons while constraining the state's ability to mount full-scale mechanized operations, as muddy terrain and low visibility limit the effectiveness of heavy vehicles and air power. Consequently, the seasonal pattern of conflict intensity may reflect the influence of climatic and agricultural cycles on both strategic opportunities and military constraints. 

\subsubsection*{d. Distance}
\begin{figure}[!h]
\centering
\includegraphics[width=1\linewidth]{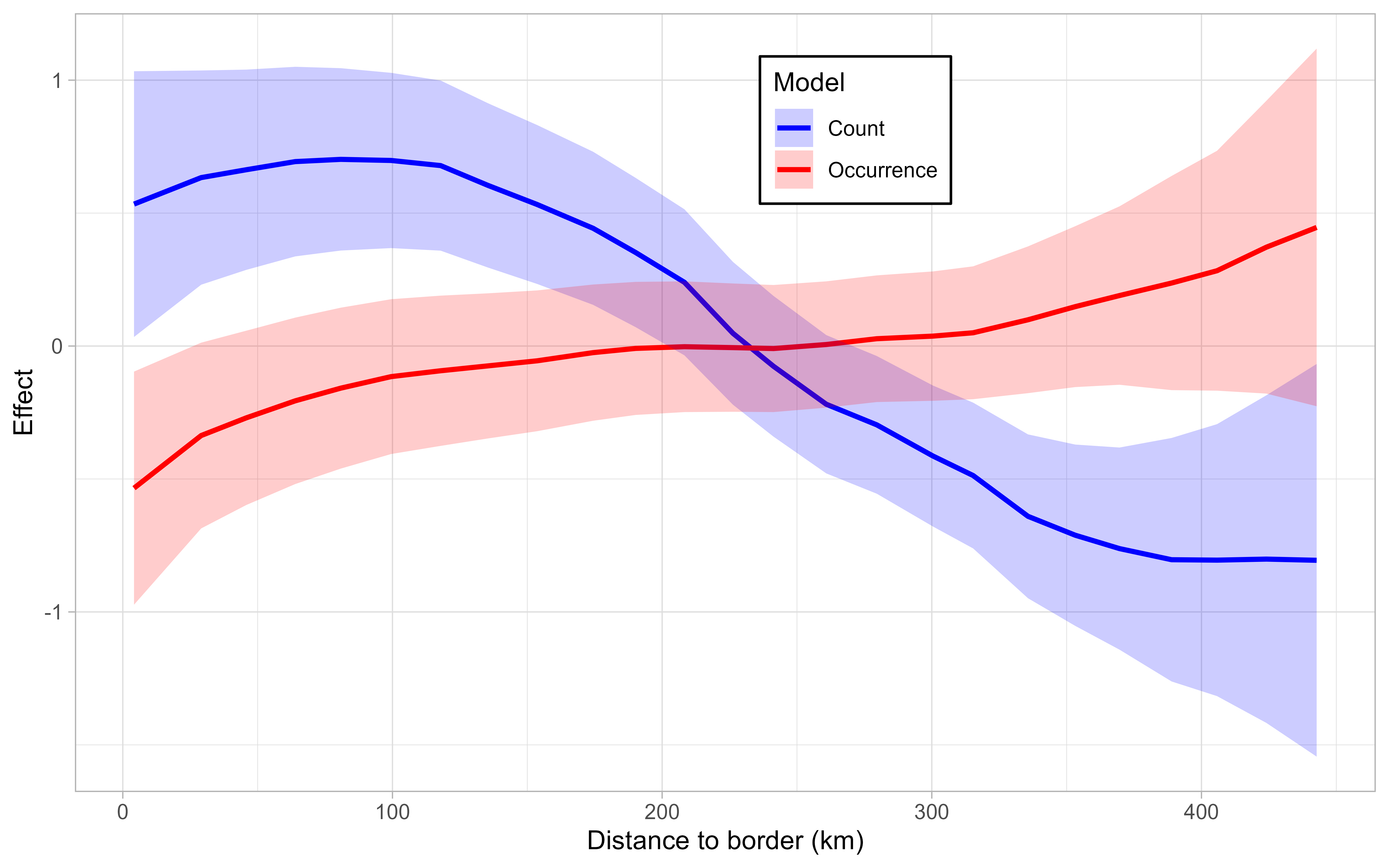}
\caption{Mean estimate and 2.5\% and 97.5\% quantiles for the posterior distribution of the nonlinear smoothed effect of distance on the linear predictor of border for the count and occurrence model.}
\label{dist_to_border}
\end{figure}

Both models included smooth functions for distance to the nearest international border and nearest major city, modeled with random walk of order two (RW(2)) priors. For distances from the border, as shown in Figure \ref{dist_to_border}, the two components show opposite patterns. 
The fatality count model indicates that events closer to borders are associated with substantially higher numbers of fatalities, with the effect gradually declining as distance increases and turning negative beyond about 200 km. In contrast, the occurrence model suggests that the likelihood of at least one fatality increases with distance from borders, becoming more pronounced beyond 200 km. This divergence suggests a key distinction: proximity to border areas is a primary driver of high-intensity, mass-casualty events. In contrast, the mere probability of a fatality occurring increases with distance from the border, but these events tend to be lower-intensity. 
In this connection, the geographical and security dynamics of Ethiopia's border areas are conducive to foreign actors (state or non-state) operating freely in adjacent territories and launching either direct or proxy warfare employing advanced military technologies. Empirical evidence supports this; for example, in the Ethiopia -- Eritrea war (1998 -- 2000) the most intense and deadliest clashes occurred in border areas \citep{dias2011conduct}.

\begin{figure}[!h]
\centering
\includegraphics[width=1\linewidth]{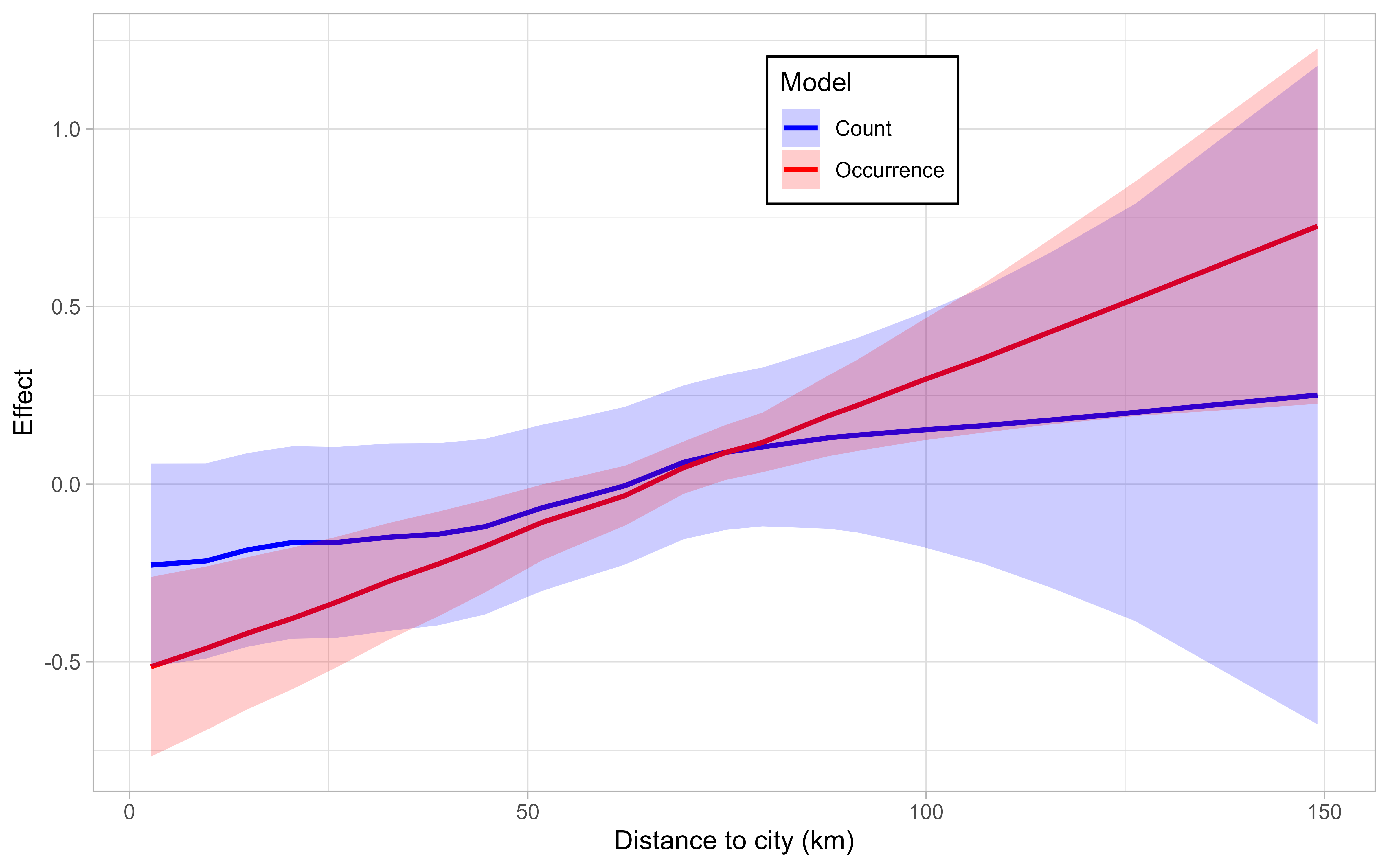}
\caption{Mean estimate and 2.5\% and 97.5\% quantiles for the posterior distribution of the nonlinear smoothed effect of distance on the linear predictor of city for the count and occurrence model.}
\label{dist_to_city}
\end{figure}

For distances from major cities, both models show broadly consistent upward trends, although the strength differs, as shown in Figure \ref{dist_to_city}. In the occurrence model, the probability of at least one fatality increases steadily with distance from urban centers. The count model shows a much more modest increase, indicating that while remoteness makes a fatal outcome more likely, it has a comparatively weaker effect on the total number of fatalities in any given event.

Taken together, these nonlinear effects suggest that border regions are hotspots for high-intensity violence, while remoteness from cities increases the likelihood that events result in fatalities, even if their intensity is lower. This is a classic pattern of insurgency or civil conflict, where state control consolidated in urban centers diminishes with distance, creating conditions for more intense and lethal armed engagements between non-state actors or between non-state and state actors. Our result thus reveals two distinct, statistically significant spatial dynamics that shape the risk of conflict fatalities in Ethiopia. Correspondingly, the decreasing frequency of conflicts in major cities reflects the strong presence of government forces and their monopoly on organized violence in regional urban centers. Overall, these results reveal two distinct, statistically significant spatial dynamics shaping the risk of conflict fatalities in Ethiopia.

\subsubsection{Temporal effects on fatalities}
\begin{figure}[!h]
\centering
\includegraphics[width=1\linewidth]{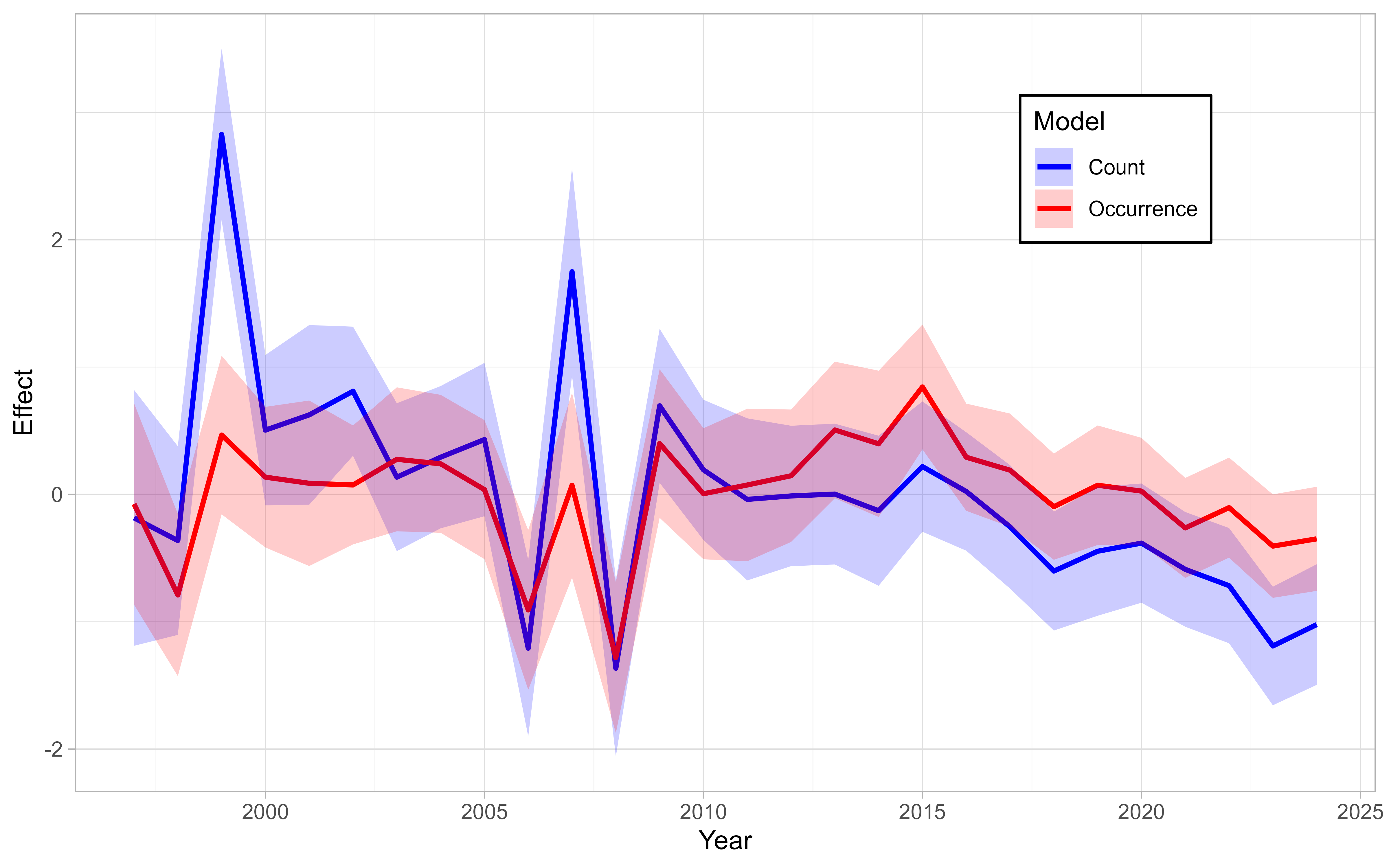}
\caption{Mean estimate and 2.5\% and 97.5\% quantiles for the posterior distribution of the year-smoothed regression effect of time (year) for the count and occurrence model.}
\label{year_random_eff}
\end{figure}

Annual variation was modeled using AR(1) priors. Figure \ref{year_random_eff} displays the estimated annual trends, revealing limited temporal persistence in both models. The occurrence model showed weak autocorrelation ($\phi_v = 0.10$ with 95\% CIs of (-0.65, 0.26)), while the count model exhibited similarly modest temporal dependence $\phi_v = 0.17$ with a 95\% CI of (-0.28, 0.43). Both credible intervals span zero, indicating substantial uncertainty in these estimates.

The temporal patterns reveal an important substantive trend. From 1997 -- 2010, conflicts were less frequent but resulted in higher fatality counts per event. After the year 2010, this pattern reversed, with more frequent conflict events happening and producing lower fatality counts. This suggests a potential shift in conflict dynamics, where either the nature of conflicts changed or reporting practices evolved.  
Overall, the weak temporal autocorrelation in both models indicates that seasonal variations and event-type characteristics explain more of the temporal pattern in conflict fatalities than smooth year-to-year dependence. 

\subsubsection{Spatiotemporal effect on fatalities}
\begin{figure}[!h]
\centering
\includegraphics[width=1\linewidth]{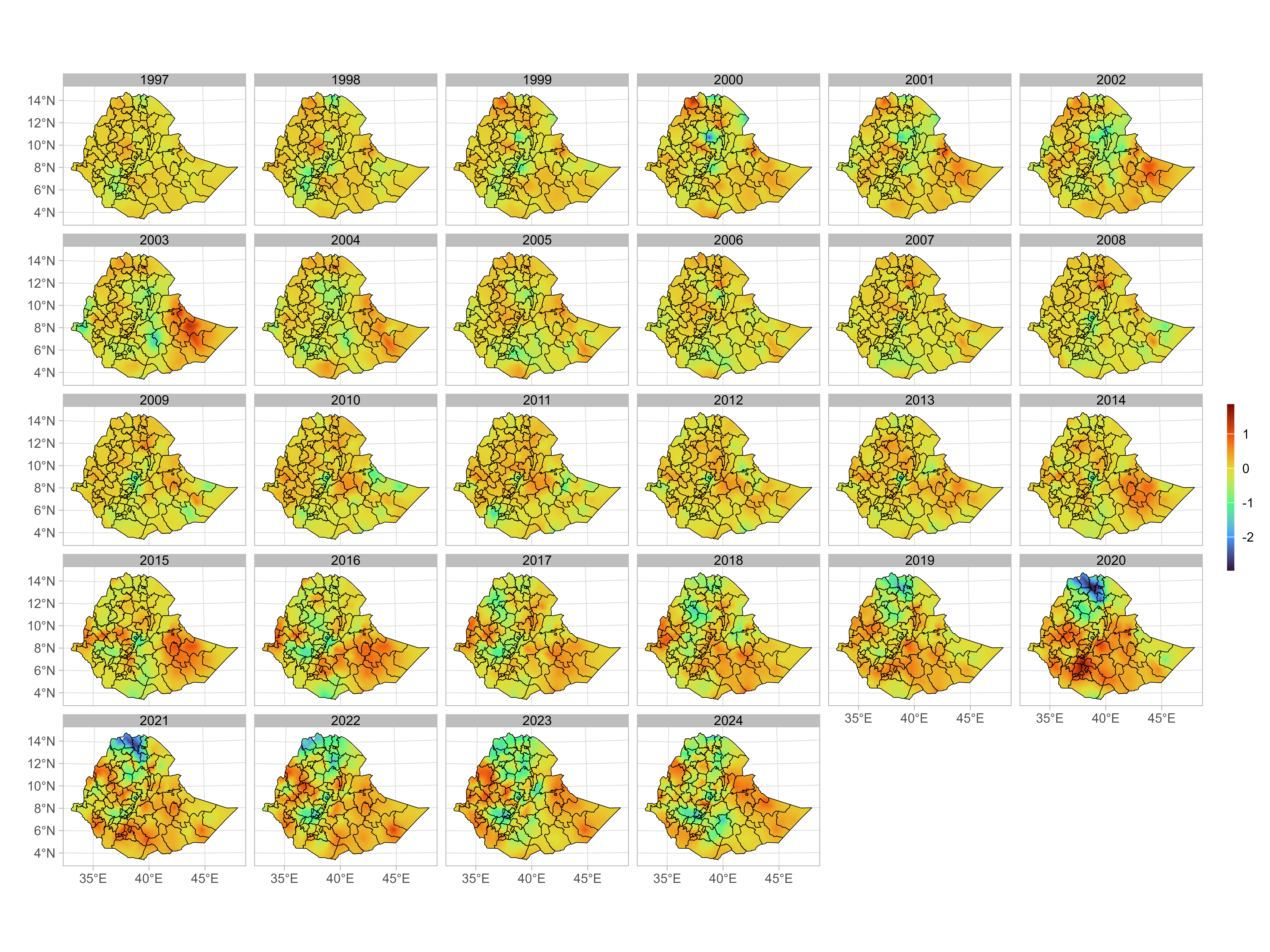}
\caption{Posterior mean of the spatiotemporal random effects of the occurrence model. The figures were generated by the authors in \texttt{R} version 4.5.1 using the ggplot2 package of version 4.0.0.}
\label{ST_random_eff_bin}
\end{figure}

\begin{figure}[h]
\centering
\includegraphics[width=1\linewidth]{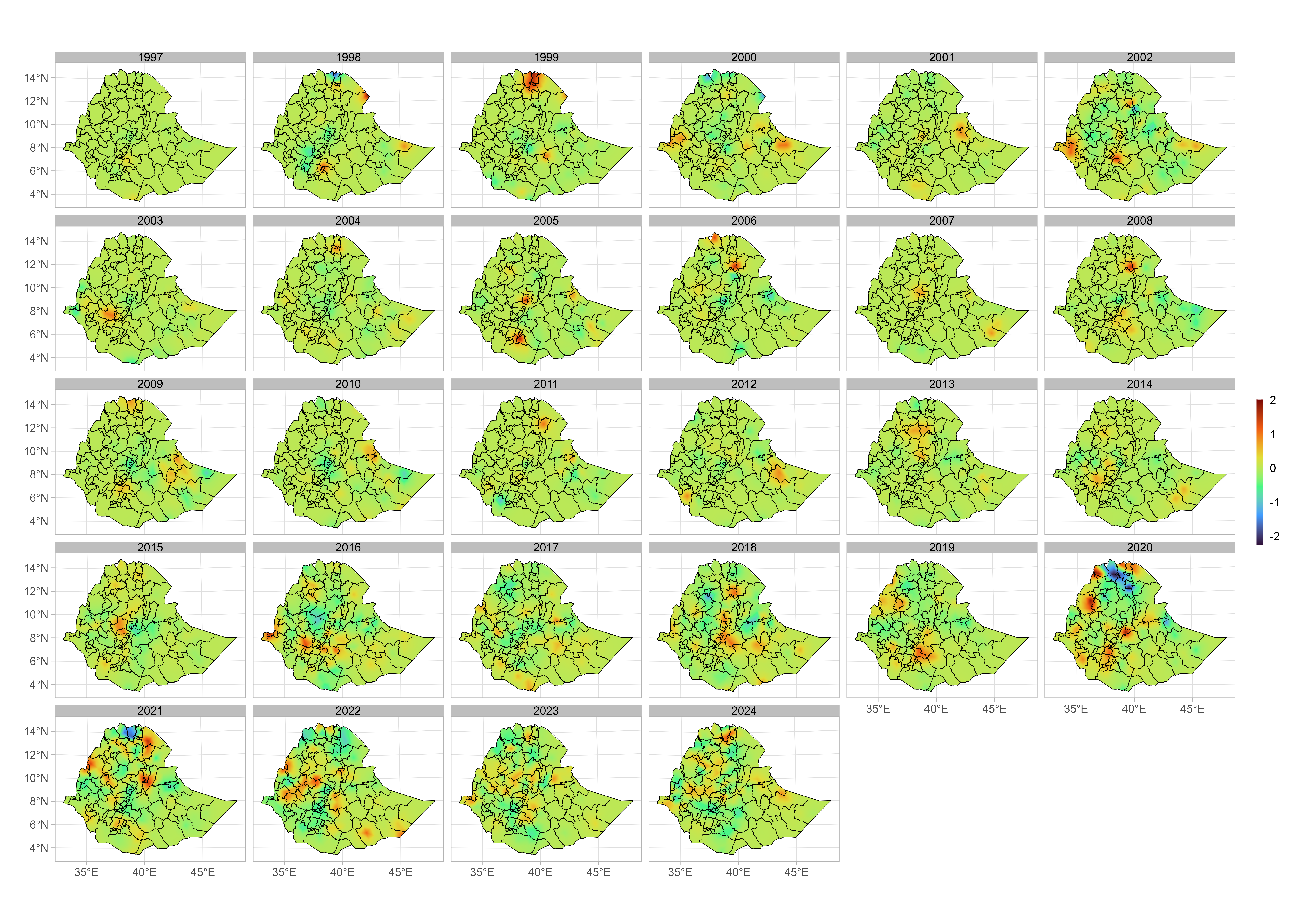}
\caption{Posterior mean of the spatiotemporal random effects of the count model. The figures were generated by the authors in \texttt{R} version 4.5.1 using the ggplot2 package of version 4.0.0.}
\label{ST_random_eff_count}
\end{figure}

The spatiotemporal field captured residual spatial clustering in both components. Figures \ref{ST_random_eff_bin} and \ref{ST_random_eff_count} present the maps of Ethiopia showing the posterior means for the spatiotemporal effect during the 28-year period for the fatality occurrence (binary component) and fatality count (count component), respectively, after adjusting for event type, season, nonlinear distance, and temporal effects. 

Figure \ref{ST_random_eff_bin} displays the estimate that represents the projected posterior mean of the estimated spatial effect on new spatial locations from the occurrence model, which is linked to the probability of fatality occurrence across the country over time through the logit function. A higher effect indicates a greater likelihood of fatality in that region, regardless of the type of violent event, season, nonlinear distances from borders and urban cities, or temporal effects. The spatiotemporal random effect for the binary occurrence model highlights non-negligible residual variation in the probability of fatality beyond that explained by observed covariates. In the late 1990s and early 2000s, the field was relatively weak, with localized positive effects; fatality occurrences due to violent events were most common in the northern and eastern border regions. From the mid-2000s, more persistent positive patterns emerge in the east and south, suggesting structural factors driving elevated fatality occurrence in these areas. Between 2016 and 2019, the field becomes more heterogeneous, with scattered positive clusters coexisting with negative effects in central Ethiopia. A marked shift is observed in 2020 -- 2021, when strong deviations occur in the north, consistent with the outbreak of the Tigray conflict. While this northern effect moderates slightly after 2022, persistent positive fields remain in western and southern areas. Overall, the estimated spatiotemporal surface captures evolving patterns of elevated occurrence risk that are not accounted for by measured covariates, underscoring the influence of unobserved and spatially structured conflict dynamics.

The projected posterior mean of the estimated spatial effects for the count model from 1997 -- 2024 is shown in Figure \ref{ST_random_eff_count}, which is directly linked to the likelihood of fatality counts through the log link function. Regions with elevated values indicate a higher number of fatalities per violent event in those areas after controlling for observed covariates. The result reveals that, during the late 1990s and early 2000s, elevated risks are concentrated along Ethiopia’s northern and eastern borders, reflecting persistent localized conflict pressures. From the mid-2000s onward, the spatial field highlights recurrent positive effects in the eastern lowlands and southern peripheries, indicating unobserved drivers of fatality clustering in these regions. Beginning around 2016, spatial contrasts intensify, with the emergence of distinct high-risk zones in western and central Ethiopia. A pronounced shift occurs in 2020 -- 2021, where the northern regions exhibit strong deviations, corresponding to the outbreak and escalation of the Tigray conflict. Although this northern effect attenuates slightly in the subsequent years, residual hotspots persist in western and southern areas through 2024. Overall, the spatiotemporal field underscores the presence of spatially structured and evolving dynamics that extend beyond measured covariates, pointing to the role of unobserved contextual and conflict-specific processes in shaping fatality counts.

\subsubsection*{Hyperparameter estimates}
The hyperparameter estimates reveal distinct spatial and temporal patterns for the two model components, as shown in Table \ref{res_hyper_para}. The spatial range was slightly shorter for the count model, approximately 114.82 km, with a 95\% credible interval (CI) of (82.75 -- 154.02), than for the occurrence model (approximately 153.13 km, with 95\% CI: 107.69 -- 209.01).  
\begin{table}[!h]
\centering
\adjustbox{max width = \textwidth}{%
\begin{tabular}{lcccccccc}
\midrule[2pt]
& \multicolumn{3}{c}{Count} &&& \multicolumn{3}{c}{Occurrence} \\
\cline{2-4} \cline{7-9}
& Mean & 2.5\% & 97.5\% &&& Mean & 2.5\% & 97.5\% \\
\midrule[1pt]
GroupRho for ST ($\phi_w$) & 0.74 & 0.65 & 0.80 &&& 0.586 & 0.44 & 0.71 \\
Range ($r$) & 114.82 & 82.75 & 154.02 &&& 153.13 & 107.69 & 209.01 \\
Stdev ($\sigma$) & 0.716 & 0.651 & 0.79 &&& 0.906 & 0.806 & 1.03 \\
Dispersion parameter (size) & 0.525 & 0.496 & 0.55 &&& - & - & - \\
Zero-inflation probability ($\psi$) & 0.622 & 0.52 & 0.74 &&& - & - & - \\
\midrule[1.25pt]
\end{tabular}}
\caption{Posterior means and 95\% credible intervals for hyperparameters.}
\label{res_hyper_para}
\end{table}
However, the occurrence model had a substantially higher marginal standard deviation as given in Table \ref{res_hyper_para}. This indicates that while the spatial influence of factors affecting fatality counts is more localized, the strength of that clustering is much stronger. In other words, the spatial drivers of high-intensity violent events are concentrated in specific, smaller hotspots, whereas the factors influencing whether any fatalities occur are more diffusely spread across a wider area.

The group-level temporal correlation was stronger for the count model ($\phi = 0.74$ with a 95\% CI of (0.65, 0.80)) than for the occurrence model ($\phi = 0.55$ with a 95\% CI of (0.44, 0.71)). This indicates that the random spatial field for fatality counts is more consistent from year to year. In other words, if a location has a high latent risk for high-fatality events in one year, that elevated risk is very likely to persist into the next year. The risk for any fatality occurring at all is less dependent on the previous year's conditions. 
Moreover, in the count model the dispersion parameter (size = 0.52) confirms substantial overdispersion in the fatality counts, justifying the use of the negative binomial distribution over a Poisson. The zero-inflation probability was estimated at $\psi = 0.622$ (95\% CI: 0.515 -- 0.74), indicating that approximately 62.2\% of the observed zero-fatality events are likely structural zeros. This substantial zero-inflation suggests that many conflict events have structural characteristics that prevent fatalities altogether. The model estimates that 37.8\% of events potentially follow the fatality-generating process, though not all of these will necessarily result in fatalities.

\subsubsection*{Posterior uncertainties}
\begin{figure}[!h]
\centering
\includegraphics[width=1\linewidth]{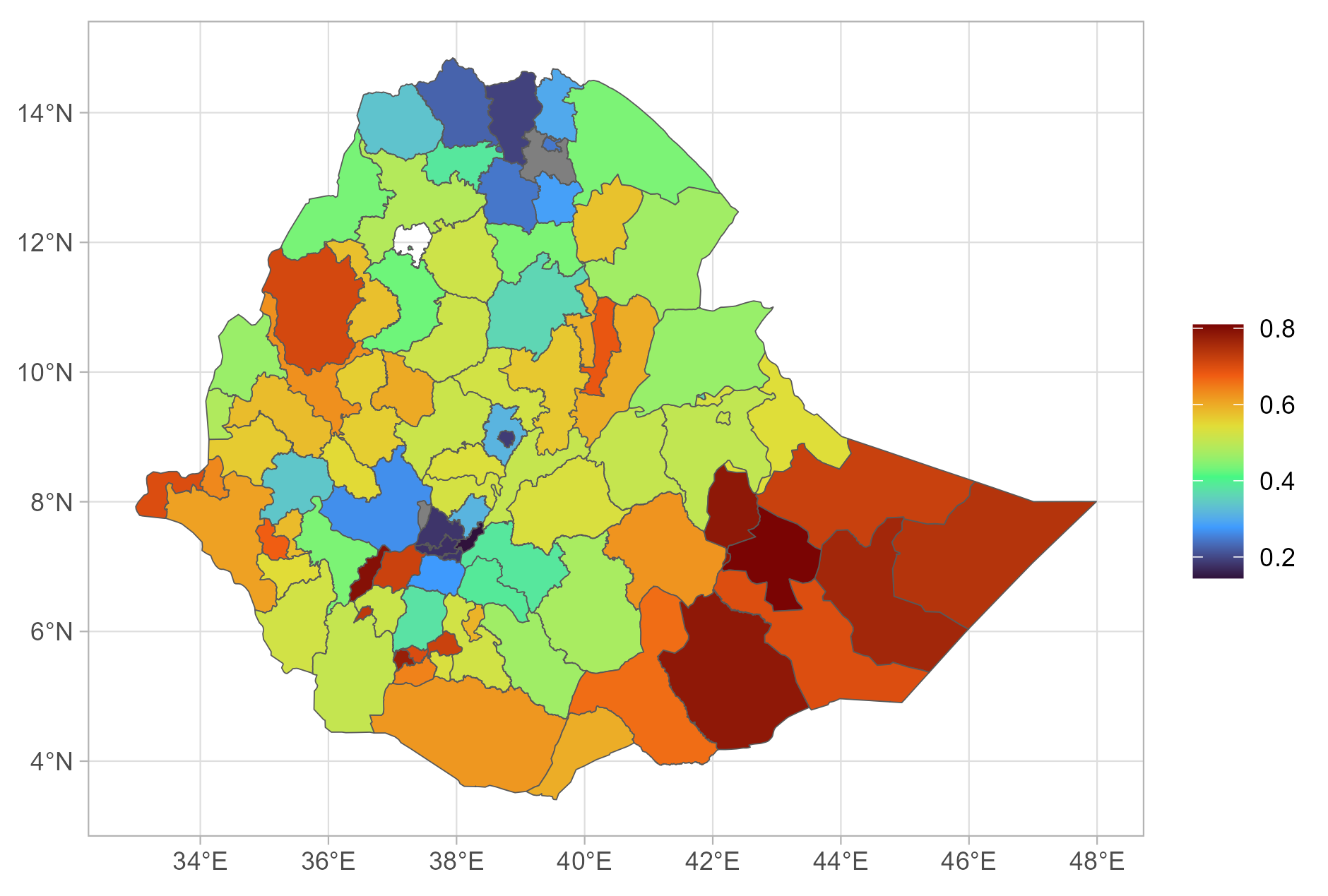}
\caption{Posterior predictive probability of fatality occurrence of violent events averaged across all years for each administrative zone of Ethiopia. The figures were generated by the authors in \texttt{R} version 4.5.1 using the ggplot2 package of version 4.0.0.}
\label{occ_prob}
\end{figure}

\begin{figure}[!h]
\centering
\includegraphics[width=1\linewidth]{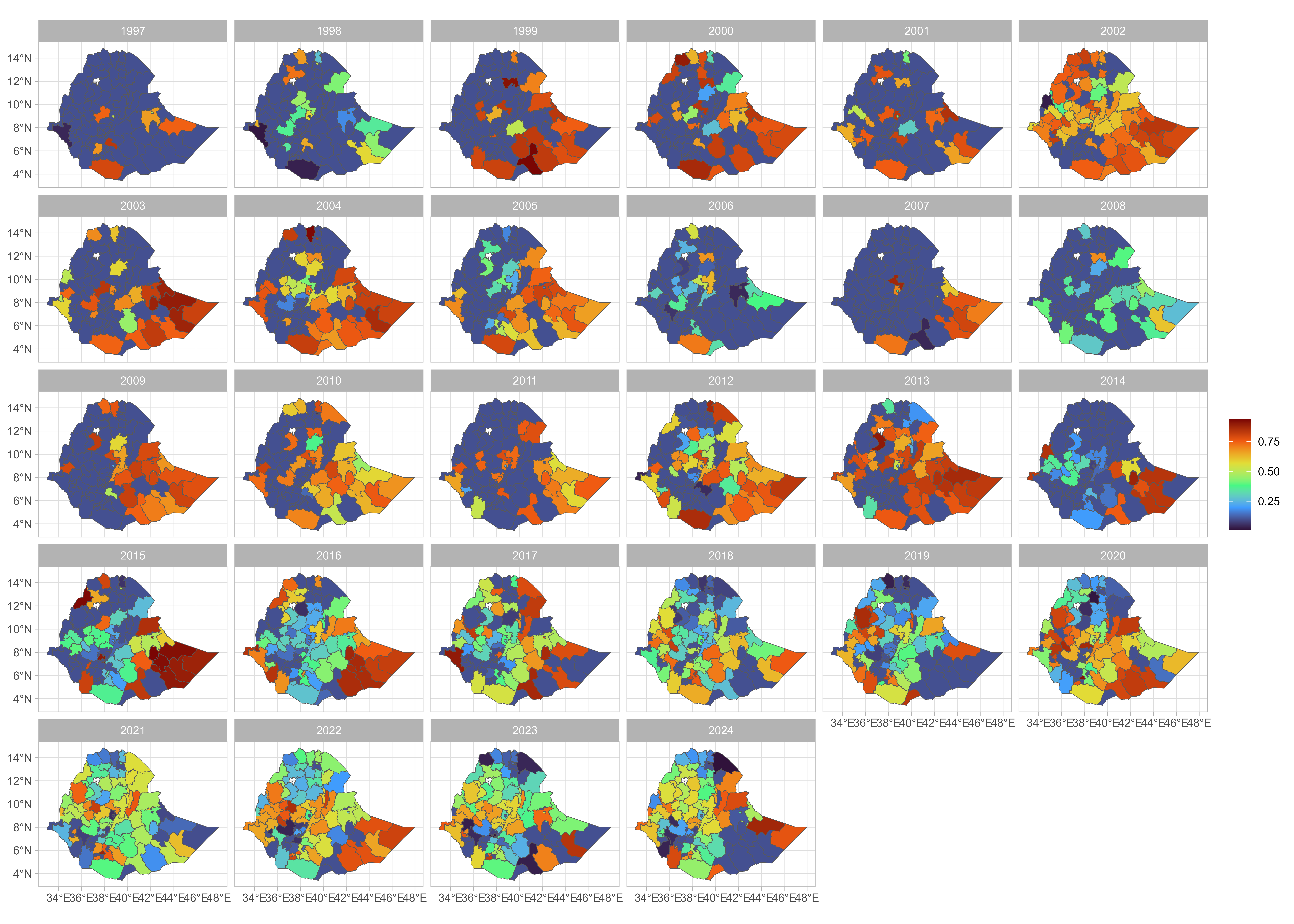}
\caption{Posterior predictive probability of fatality occurrence for each year from 1997 to 2024 for each administrative zone of Ethiopia. The figures were generated by the authors in \texttt{R} version 4.5.1 using the ggplot2 package of version 4.0.0.}
\label{ZoneYear_occ_prob}
\end{figure}

Figure \ref{occ_prob} shows the posterior probabilities of the fatality occurrence due to violent events aggregated for each administrative county (zone) to ease comparison, integrating over all the years considered in this study. 
The higher the probability, the more likely an area would experience higher fatality. The result showed that fatality is most likely to occur in the west and northwest, south and southeast, and northeast-central Ethiopia.
Moreover, Figure \ref{ZoneYear_occ_prob} shows the predictive probability of fatality occurrence of violent events for each year from 1997 to 2024 for each administrative zone, for ease of comparison of the spatiotemporal changes over Ethiopia. 

%%%%%%%%%%%%%%%%%%%%%%%%%%%%
%%%%%%%%%%%%%%%%%%%%%%%%%%%%   %Discussion
%%%%%%%%%%%%%%%%%%%%%%%%%%%%

\section{Discussion}\label{sec_discussion}
This study presents a fully Bayesian spatiotemporal analysis of conflict fatalities and occurrences due to violent events in Ethiopia from 1997 to 2024. We employed dual modeling approaches to disentangle the drivers of fatality occurrence and intensity. By integrating a zero-inflated negative binomial model for fatality counts with a Bernoulli model for occurrence probability, we addressed both dimensions of conflict lethality while accounting for population exposure, spatial dependence, and temporal correlation. Our framework incorporated categorical predictors (event types, actors, season), nonlinear distance effects, and structured random effects to capture spatiotemporal heterogeneity. This approach revealed that conflict fatality is not binary but exists on a continuum, with distinct factors influencing whether fatalities occur versus how many occur when violent events happen.
%

%% Event types
The finding reveals that violent events demonstrate differential impacts on fatality dimensions in Ethiopia. Air/drone strikes, shelling/artillery attacks, and conventional attacks consistently increased both fatality occurrence probability and count (or intensity), representing comprehensive escalation events. The particularly strong effect of air/drone strikes (occurrence: odd ratio (OR) = 3.15, count: incidence rate ratio (IRR) = 2.25) suggests these events create conditions where fatalities are not only more likely to occur but also more severe when they do. This pattern aligns with findings from \citet{EgbonGayawan2025}, where aerial attacks were associated with high collateral damage due to questionable accuracy in targeting and civilian distribution amidst conflict zones. Conversely, sexual violence, strategic developments, excessive force against protesters, and grenade attacks consistently reduced fatality across both dimensions. The protective effect of sexual violence events, while counterintuitive, may reflect underreporting, different conflict modalities, or the fact that these events often occur without immediate lethal outcomes despite their severe humanitarian consequences. This finding contrasts with the Nigerian context, where sexual violence showed high fatality associations, suggesting contextual differences in how sexual violence manifests within conflict settings \cite{EgbonGayawan2025}. The dimension-specific effects are particularly revealing. Remote explosions significantly reduced fatality counts but not occurrence probability, suggesting these events may cause injuries rather than deaths when they do result in casualties. Territory overtakes reduced occurrence probability but not fatality intensity, indicating that when these strategically significant events turn lethal, they can be particularly severe.

%% Actors
Findings revealed distinct mechanistic pathways through which different actors influence conflict fatality occurrences and intensities. Foreign actors demonstrate a significant increase in conflict fatalities, quadrupling the fatality counts and greatly enhancing the probability of occurrence. This aligns with theories of conflict internationalization, where external involvement may introduce advanced weaponry, different tactical doctrines, or reduced accountability mechanisms \citep{patricia2019strategies, GALLEA2023103001}. Specifically, Ethiopia has experienced cross-border conflicts involving state and non-state actors from neighboring countries, often leveraging ethnic kinship with communities across Ethiopia's borders to conduct high-intensity operations. Communal groups show a balanced moderate effect (40\% increase in both models), reflecting their deep local embeddedness and the sustained nature of ethnic and identity-based conflicts that characterize much of Ethiopia's violence landscape. This pattern reflects the characteristics of ethnic conflicts in Ethiopia, where historical grievances, territorial disputes, and identity politics create conditions for both frequent outbreaks and sustained intensity. Rebel forces and militia groups exhibit a different pattern, increasing fatality intensity without affecting occurrence probability. This suggests these actors may influence conflict intensity rather than the basic likelihood of lethal outcomes. In contrast, civilians, protesters, and rioters show negative effects across both models, indicating fundamentally different conflict modalities with inherent fatality constraints.
%

%% Season
This study revealed notable seasonal variations in the incidence of fatalities resulting from violent events in Ethiopia. The results indicate that fatalities were significantly higher during the summer season compared to spring. This discovery aligns with and is thoroughly explained by the complex framework of \citet{Raleigh2012conflictandclimate}. Conventional narratives often link resource scarcity in the dry season to increased conflict; however, our findings, consistent with their disaggregated analysis, reveal that the rainy summer is a time of heightened lethal risk. The mechanism appears to consist of two components: \citet{Raleigh2012conflictandclimate} demonstrating that in East Africa, rebel conflict intensifies during periods of atypical drought, while communal violence escalates during periods of atypical rainfall. During the summer in Ethiopia, these conditions frequently occur simultaneously in various locations. The western highlands experience the most substantial rainfall, potentially inciting conflicts over arable land and livestock, whereas the eastern lowlands remain arid, facilitating the mobility of insurgent factions. The overall increase in fatalities during the summer is likely attributable to the convergence of these two distinct types of conflict, each exacerbated by varying weather conditions across the country. This explains why the most lethal occurrences, such as organized campaigns and significant confrontations between groups, are more prevalent during this season.

%% Distances
Distance from international borders and cities shows distinct patterns. The result revealed that events near borders are more intense, with higher fatality counts, while those farther inland are more likely to turn lethal even if less severe. Similarly, remoteness from cities increases the probability of a fatal event, though it has only a minor effect on the number of fatalities. This is supported by the reports from ACLED\footnote{\url{https://acleddata.com/update/violence-patterns-ethiopias-periphery-march-2024}} and studies by \citet{stefan2024spatialpattern} and \citet{EID2024e38684}. These results suggest that border regions concentrate high-intensity violence, whereas more remote areas face a higher risk of fatality occurrence. Taken together, these results reveal two distinct spatial dynamics of conflict risk: (1) a border dynamic governing the intensity of violence and (2) an urbanicity dynamic governing the probability of violence. This is a classic pattern of conflict insurgency \citep{Tollefsen2015Insurgency}, where state control or monopoly of violence diminishes with distance from urban centers, creating conditions for persistent, lower-intensity violence. Meanwhile, along the international territorial borders with neighboring countries are often hotspots for strategic, high-intensity engagements between armed actors due to historically rooted geopolitical and geostrategic vulnerabilities of the country to neighboring countries and regional and global powers, which have long-term interests in countries of the Horn of Africa, notably Ethiopia. Thus, the study identifies two significant dynamics shaping conflict fatalities in Ethiopia.

%Note: Conflicts may cluster very close to borders, decline in mid-range, then rise again near large cross-border routes. For cities, risk may peak in peri-urban areas but be lower inside dense city centers

%% spatiotemporal
Spatial analysis identified heterogeneous risk patterns across Ethiopia, with certain regions maintaining elevated fatality risks throughout the study period. The SPDE approach effectively captured continuous spatial variation, revealing that conflict hotspots often transcend administrative boundaries. The growing intensity of conflicts along these boundaries is linked to the politicization of ethnicity, a key feature of Ethiopia's post-1991 federal system, which has generated competing interests among regional (or zonal) governments over contested territories and produced complex geographical patterns. Findings from the result showed significant heterogeneity in fatality due to violent events across years and locations in the country. Overall, violence has evolved through time, and the fatalities resulting from these events vary significantly across the country. The result obtained identified places in the northern, northeastern, western, southwestern, central, south-central, and near-the-border regions as the most exposed regions to fatality in Ethiopia.

%%Limitations and future research
Although the model employed in this study is robust, it does have certain limitations. One key limitation is the use of an autoregressive model of order one (AR(1)) to model temporal dependencies within the proposed spatiotemporal framework. While this approach effectively captures short-term temporal dependence, it fails to account for more complex dependencies that data-driven order selection could better represent. Future work could investigate alternative specifications to enhance model flexibility and interpretability in capturing diverse temporal patterns.

Another limitation is that we analyzed fatality counts and occurrences separately, whereas INLA currently does not allow fully flexible, separate linear predictors for the zero-inflation probability, instead using a single predictor. Although we employed the occurrence model for validation purposes, future research could explore a spatiotemporal hurdle model for both fatality and occurrence, incorporating shared effects, which INLA can accommodate. Additionally, a Bayesian spatial joint model combining a Bernoulli component for occurrence and a Poisson component for counts could address conflict fatality mapping with excessive zeros, incorporate shared spatial effects, and leverage the areal model with penalized complexity priors \cite{asmarian2019bayesian}. Another concern is potential data reporting bias, specifically on the fatality count. Despite using mesh triangulation, the analysis may not fully correct biases related to the reporting of violent event locations. Future work could develop more robust methods to account for such biases.

Furthermore, this study assumes that event locations are non-random. Future research could tackle this assumption by utilizing models that account for randomness in location, such as point process models, specifically log-Gaussian Cox process models. Incorporating economic, socioeconomic, and demographic variables into the modeling framework could also enhance our understanding of their influence on fatalities arising from violent events. Lastly, given that event types exhibit different scales of fatality, future research should focus on developing multilevel point processes with shared effects to capture complex conflict dynamics. This approach could improve both predictive accuracy and mechanistic understanding.

\section{Conclusion} \label{sec_conclusion}
This study applied a Bayesian hierarchical spatiotemporal framework to analyze conflict fatalities in Ethiopia, advancing understanding of both their occurrence and intensity. The results reveal a distinct North–South divide, with northern regions more exposed to lethal events, and demonstrate clear seasonal variation in fatality risks. Event types and actor identities emerged as critical determinants: airstrikes, attacks, and shelling generated higher fatality counts in reference to armed clashes, while foreign and communal (identity) groups affected both the occurrence and count of fatalities, and rebel and militia groups shaped the severity of lethal outcomes without strongly affecting their likelihood. Conversely, protests and riots showed consistent protective effects, reflecting different conflict dynamics with lower inherent lethality.

By modeling occurrence and intensity as separate processes, this analysis provides more nuanced insights than single-model approaches and highlights the prevalence of non-lethal events, suggesting opportunities for early intervention. The resulting spatiotemporal risk maps offer practical tools for humanitarian planning, allowing evidence-based allocation of resources and more targeted conflict mitigation strategies. Together, these findings underscore the value of integrating methodological innovation with policy relevance in studying conflict fatality, offering a foundation for more effective response in Ethiopia’s evolving security landscape.

%%%%%%%%%%%%%%%%%%%%%%%%%%%%
%%%%%%%%%%%%%%%%%%%%%%%%%%%%   %Acknowledgments
%%%%%%%%%%%%%%%%%%%%%%%%%%%%

\paragraph{Acknowledgments:} Yassin Tesfaw Abebe acknowledges support from the International Mathematical Union (IMU) and the 	Graduate Research Assistantships in Developing Countries (GRAID) Program. We would like to express our sincere appreciation to the Armed Conflict Location and Event Data (ACLED) project for kindly providing the necessary data for our research.
\paragraph{Conflict of interest:} The authors declare that they have no conflict of interest.
%\paragraph{Research funding:} No funding was received.

%%%%%%%%%%%%%%%%%%%%%%%%%%%%
%%%%%%%%%%%%%%%%%%%%%%%%%%%%   %References
%%%%%%%%%%%%%%%%%%%%%%%%%%%%

\bibliographystyle{plainnat}
\bibliography{refs}

%%%%%%%%%%%%%%%%%%%%%%%%%%%%
%%%%%%%%%%%%%%%%%%%%%%%%%%%%   %Appendix
%%%%%%%%%%%%%%%%%%%%%%%%%%%%

\appendix
\section*{Appendix}
In this section, we give further theoretical details of the temporal, spatiotemporal, and nonlinear models used in this work as given in Eq. \eqref{lin_pred}. In addition, a detailed specific prior is used for each hyperparameter.

\subsection*{A.1 Specification of the nonlinear distance effect}
The nonlinear distance effect of distances from the event location to the nearest international border is modeled using a random walk (RW2) prior. The function $f_b(d_i)$ is defined discretely over a set of $m$ ordered distance points (knots) defined on conflict event location $s_i$, $\mathbf{d} = (d_1, \dots, d_n)^\top$. Let $\mathbf{f} = (f_1, f_2, \dots, f_n)^\top$ denote the values of the smooth function at these points, i.e., $f_i = f_b(d_i)$. The RW2 prior is defined through its structure on the second-order differences of the sequence $\mathbf{f}$, asserting that the function is locally quadratic and deviations from this are penalized. The prior is specified by
\begin{equation}
	\Delta^2 f_i = f_i - 2f_{i-1} + f_{i-2} \sim \mathcal{N}(0, \tau^{-1}) \quad \text{for } i = 3, 4, \dots, n
\end{equation}
where $\tau$ is a precision hyperparameter controlling the smoothness of the function. Higher values of $\tau$ enforce a smoother function. This conditional specification implies a multivariate Gaussian prior for the entire vector $\mathbf{f}$ as
\begin{equation}
	\mathbf{f} \mid \tau \sim \mathcal{N}(\mathbf{0}, (\tau \mathbf{R})^{-1}).
\end{equation}

The precision matrix $\mathbf{Q} = \tau \mathbf{R}$ is sparse and structured. The matrix $\mathbf{R}$ is defined such that $\mathbf{f}^\top \mathbf{R} \mathbf{f} = \sum_{i=3}^n (\Delta^2 f_i)^2$, it penalizes sharp changes in the second derivative. To ensure an identifiable model, the function is constrained to have a zero mean, $\sum_{j=1}^n f_i = 0$. A Penalized Complexity (PC) \cite{simpson2017penalising} prior was placed on the standard deviation $\sigma = 1 / \sqrt{\tau}$ of the random effect. The PC prior is designed to penalize departure from a simpler base model (in this case, $\sigma = 0$, implying a constant function $f(d) = 0$). It is defined by
\begin{equation*}
	P(\sigma > U) = \alpha.
\end{equation*}

In our analysis, we set $U = 0.5$ and $\alpha = 0.01$, specifying our prior belief that there is only a 1\% probability the standard deviation of the smooth effect exceeds 0.5 on the log-risk scale. This corresponds to the following prior density on the precision
\begin{equation*}
	p(\tau) = \frac{\lambda}{2} \tau^{-3/2} \exp \left( -\lambda \tau^{-1/2} \right), \quad \text{where } \lambda = -\frac{\ln(\alpha)}{U}.
\end{equation*}

Within the Integrated Nested Laplace Approximation (INLA) framework, the posterior marginals for the function values $\mathbf{f}$ and the hyperparameter $\tau$ are approximated. The resulting posterior mean, $E[f_j \mid \mathbf{y}]$, and credible intervals for each $f_j$ are plotted to visualize the estimated nonlinear relationship between distance to the border and the log-risk of conflict fatalities, as shown in Figure \ref{dist_to_border} in the main text.

\subsection*{A.2 Specification of temporal random effect}
The temporal random effect $v_t$, for the year is modeled using a first-order autoregressive process (AR(1)) to capture potential correlation between consecutive years. Let $\mathbf{v} = (v_1, v_2, \dots, v_T)^\top$ represent the vector of temporal random effects for $T$ ordered years. The AR(1) process defines the value at time $t$ conditional on the previous value is
\begin{equation}
	v_t = \phi_v v_{t-1} + \epsilon_t, \quad \text{for } t = 2, 3, \dots, T
\end{equation}
where $|\phi_v| < 1$ is the temporal autocorrelation parameter, $\epsilon_t \sim \mathcal{N}(0, \tau^{-1})$ are independent innovations, and $\tau$ is the precision of the innovation term. This implies a multivariate Gaussian prior for $\mathbf{v}$ given by
\begin{equation}
	\mathbf{v} \mid \phi_v, \tau \sim \mathcal{N}(\mathbf{0}, (\tau \mathbf{Q}(\phi_v))^{-1}),
\end{equation}
where the precision matrix $\mathbf{Q}(\phi)$ for the AR(1) process is tridiagonal and given by
\begin{equation*}
	\mathbf{Q}(\phi_v) = 
	\begin{bmatrix}
		1 & -\phi_v & 0 & \cdots & 0 \\
		-\phi_v & 1+\phi_v^2 & -\phi_v & \ddots & \vdots \\
		0 & -\phi_v & 1+\phi_v^2 & \ddots & 0 \\
		\vdots & \ddots & \ddots & \ddots & -\phi_v \\
		0 & \cdots & 0 & -\phi_v & 1
	\end{bmatrix}.
\end{equation*}
Similarly, we use PC priors \cite{fuglstad2019constructing} for both hyperparameters: for precision $\tau$: $P(1/\sqrt{\tau} > U_\sigma) = \alpha_\sigma$ and for the autocorrelation $\rho$: $P(\rho > U_\rho) = \alpha_\rho$. In the analysis, we set $\pi(\sigma > 0.5) = 0.01$ where $\sigma = 1/\sqrt{\tau}$ and $\pi(\rho > 0) = 0.9$ to reflect a prior belief in positive temporal correlation.

%To ensure identifiability with the model intercept, the vector $\mathbf{v}$ is constrained to sum to zero ($\sum_{t=1}^T v_t = 0$).

\subsection*{A.3 Spatiotemporal random effects specification}
The spatiotemporal dependence structure for the conflict event fatalities is modeled by combining a Gaussian random field (GRF) for spatial effects with a first-order autoregressive process for temporal dynamics. The following section details the mathematical specification of this structure, which is implemented using the stochastic partial differential equation (SPDE) approach in R-INLA \citep{lindgren2011explicit}. We model spatial dependence using GRFs with Mat\'ern covariance functions, approximated through the SPDE approach. This yields a sparse Gaussian Markov random field (GMRF) \citep{rue2005gaussian} representation defined on a triangulated mesh. Temporal dependence is introduced via a first-order autoregression (AR(1)) applied to the latent spatiotemporal field $w(s,t)$. The interaction evolves as
\begin{equation}\label{eq_spde_ar1}
	w(s,t) = \phi_w  w(s,t-1) + \omega(s,t), \qquad |\phi_w| < 1,
\end{equation}
where $\phi_w$ is the AR(1) coefficient and $\omega(\cdot,t)$ is a mean-zero spatial GRF with Mat\'ern covariance. For two conflict event locations $s_i,s_j$ at distance $h=\|s_i-s_j\|$,
\begin{equation}\label{eq_matern}
	\mathrm{Cov}\!\big(\omega(s_i,t),\omega(s_j,t)\big)
	=\frac{\sigma_\omega^2}{\Gamma(\nu) 2^{\nu-1}}(\kappa h)^\nu K_\nu(\kappa h),
\end{equation}
with marginal variance $\sigma_\omega^2$, smoothness $\nu>0$, scale $\kappa>0$, and modified Bessel function $K_\nu(\cdot)$. The correlation range is $\rho=\sqrt{8\nu}/\kappa$.

\begin{figure}[!h]
	\centering
	\includegraphics[width=1\linewidth]{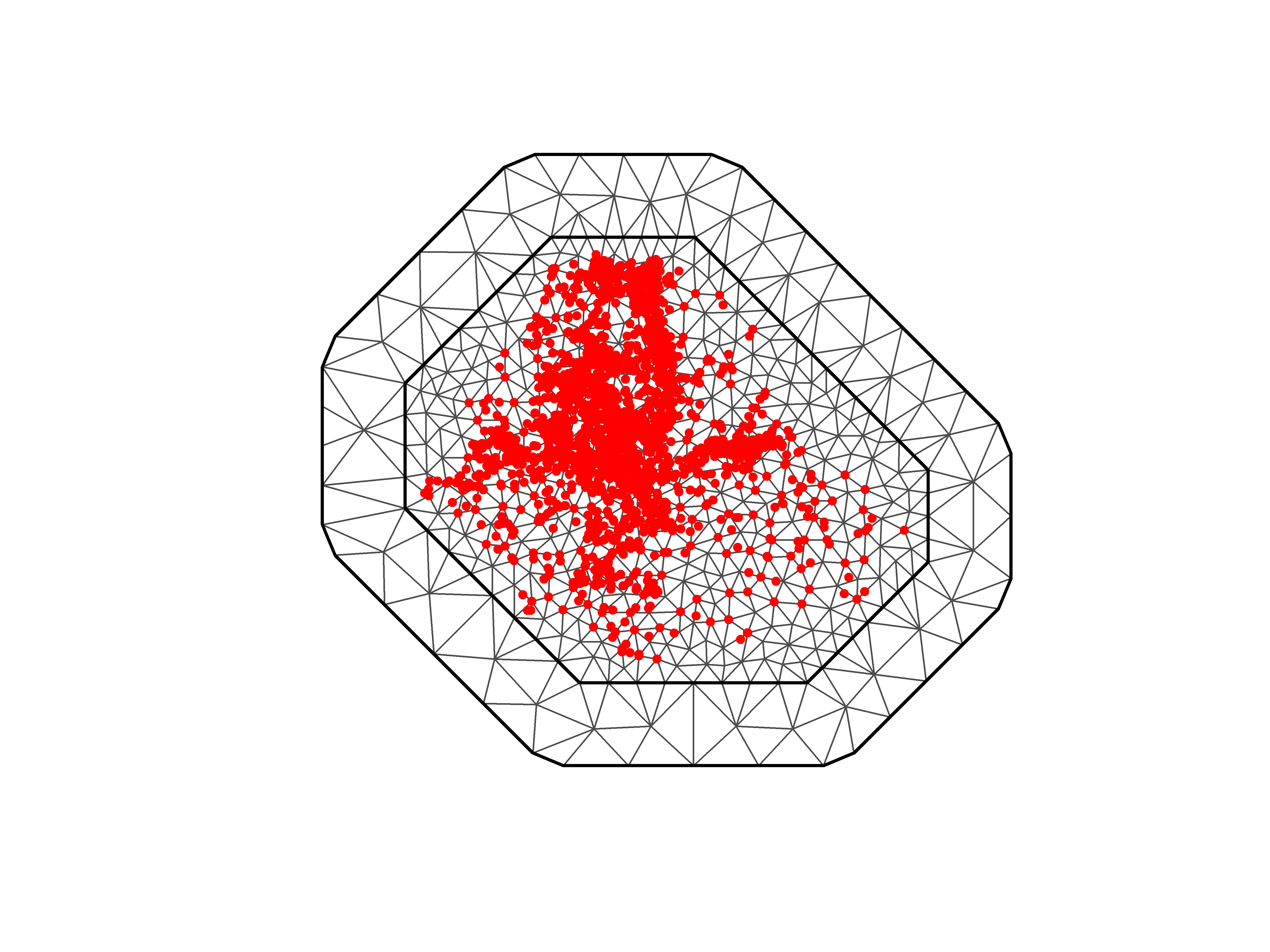}
	\caption{Constructed triangular mesh for the study region obtained using a finite element method. The red dots denote observed event locations from 1997 to 2024.}
	\label{mesh}
\end{figure}

The field $\omega(\cdot,t)$ satisfies the SPDE
\begin{equation}\label{eq_spde}
	(\kappa^2-\Delta)^{\alpha/2} (\tau \,\omega(s,t)) = \mathcal{W}(s,t),
\end{equation}
where $\Delta$ is the Laplacian, $\mathcal{W}(s,t)$ denotes Gaussian white noise, $\alpha$ satisfies $\nu=\alpha-d/2$ in $d=2$ dimensions, and $\tau>0$ controls marginal variance ($\sigma_\omega^2 \approx 1/(4\pi \tau^2 \kappa^{2\nu})$). Discretization on a Delaunay mesh (Figure \ref{mesh}) with $m$ vertices gives
\begin{equation*}
	\omega(s,t) \ \approx \ \sum_{k=1}^{m}\psi_k(s)\ \theta_k(t),
\end{equation*}
with basis functions $\{\psi_k\}$ and coefficients $\boldsymbol{\theta}(t)\sim\mathcal{N}(0,\mathbf{Q}_\omega^{-1})$, where $\mathbf{Q}_\omega$ is the sparse SPDE precision matrix. Let $\mathbf{A}_s\in\mathbb{R}^{n\times m}$ project the mesh basis to $n$ observed sites, with entries $A_{ik}=\psi_k(s_i)$. Then
\begin{equation}\label{eq_projection}
	\boldsymbol{\omega}(t) = (\omega(s_1,t),\ldots,\omega(s_n,t))^\top 
	\approx \mathbf{A}_s \boldsymbol{\theta}(t).
\end{equation}

Under the assumption of a separable spatiotemporal structure, the joint precision matrix for the entire discretized field $\mathbf{w} = (\boldsymbol{\theta}(1)^\top,\dots,\boldsymbol{\theta}(T)^\top)^\top$ is given by the Kronecker product as
\begin{equation}\label{eq_block}
	\mathbf{Q}_{st} = \mathbf{Q}_{\text{AR(1)}}(\phi_w)\otimes\mathbf{Q}_\omega,
\end{equation}
where $\mathbf{Q}_{\text{AR(1)}}(\phi_w)$ is the $T\times T$ AR(1) precision matrix, $\mathbf{Q}_\omega$ is the $m\times m$ spatial precision matrix, and $\otimes$ is the Kronecker product. This structure ensures computational scalability, and to ensure model identifiability with the global intercept, the spatiotemporal field $\mathbf{w}$ was constrained to sum to zero over the domain for each time point \cite{rue2009approximate}. 

\end{document}